\documentclass[%
 reprint,
 amsmath,amssymb,
 aps,
]{revtex4-2}

\usepackage{graphicx}
\usepackage{dcolumn}
\usepackage{bm}
\usepackage{xcolor,soul,framed} 
\usepackage{subfigure}
\usepackage{fancyhdr}
\usepackage{booktabs}
\usepackage{tabularx}

\newcommand{\Fig}[1]{Figure\,{\ref{#1}}}

\begin{document}

\title{{\footnotesize  Accepted for publication in Physical Review Applied. doi: https://doi.org/10.1103/PhysRevApplied.19.044014} 
{\\Diffraction Phenomena in Time-varying Metal-based Metasurfaces}}

\

\author{Antonio Alex-Amor$^{*}$}
\affiliation{
Department of Information Technology, 
Universidad San Pablo-CEU, CEU Universities,  Campus Montepríncipe, 28668 Boadilla del Monte (Madrid), Spain
}%
\author{Salvador Moreno-Rodríguez, Pablo Padilla, Juan F. Valenzuela-Valdés, Carlos Molero}
\affiliation{%
Department of Signal Theory, Telematics and Communications, Universidad de Granada, 18071 Granada, Spain
}%

\begin{abstract}
This paper presents an analytical framework for the analysis of time-varying {metal-based} metamaterials. Concretely, we particularize the study to time-modulated {metal-air} interfaces embedded between two different semi-infinite media that are illuminated by monochromatic plane waves of frequency $\omega_0$. The formulation is based on a Floquet-Bloch modal expansion, which takes into account the time periodicity of the structure ($T_s = 2\pi / \omega_s)$, and integral-equation techniques. It allows to extract the reflection/transmission coefficients as well as to derive nontrivial features about the dynamic response and dispersion curves of time-modulated {metal-based} screens. In addition, the proposed formulation has an associated analytical equivalent circuit that gives physical insight to the diffraction phenomenon. Similarities and differences between space- and time-modulated metamaterials are discussed via the proposed circuit model. Finally, some analytical  results are presented to validate the present framework. A good agreement is observed with numerical computations provided by a self-implemented finite-difference time-domain (FDTD) method. Interestingly, the present results suggest that time-modulated {metal-based} screens can be used as pulsed sources (when $\omega_s \ll \omega_0$), beamformers 
($\omega_s \sim \omega_0$) to redirect energy in specific regions of space, and analog samplers ($\omega_s \gg \omega_0$).
\end{abstract}

\maketitle


\section{\label{sec:introduction}Introduction}

The propagation properties of electromagnetic waves through time-varying media is a topic that has classically been studied to understand effective modulations produced by interaction of two waves \cite{Tamir64, Elachi1971}. In the last years the topic has been revisited due to the potential applications of time-varying systems in communications. Exotic properties related to the inherent non-reciprocity \cite{Taravati2019, Shaltout2015} or beam-steering capabilities \cite{Zhang2018} attracted the interest of many researchers. At the same time, the theoretical background associated with time-varying systems has benefited a fruitful development. This is the case of the generalization of Kramer-Kronig relations to temporal scenarios \cite{Solis2021}, dual behaviors in spacially- and temporally-varying systems \cite{Pacheco2021}, or the discovery of promising scenarios  based on moving gratings \cite{Pendry2022}. A complete compilation of spacetime media is reported in \cite{Caloz2020_1, Caloz2020_2}, exposing the general concepts and examining theoretical implications and promising application fields. 

Modern time-varying systems are based on periodic-structure and metasurface configurations, whose architecture can approximately be interpreted as periodic or quasi-periodic distributions of individual emitters or meta atoms \cite{Tiukuvaara2021, Ptitcyn2019}. Following this line of argument, further definitions are discussed for temporal systems in \cite{Yin2022review}. An original concept arises by mixing classical metamaterials and metasurfaces with time modulation. The fundamental advantage of this conception is the long tradition of metamaterials/metasurfaces in the microwave and photonics communities, and the vast knowledge accumulated in the last 20 years from both the experimental and manufacturing point of view \cite{Marques2008, Munk2009, Alex3D_2022}. Time modulation can individually be  incorporated on the meta atoms by reconfigurable elements, such as mechanical, electrical or optical, among others \cite{Shaltout2019}.    

These spatiotemporal metamaterials/metasurfaces have succeeded in performing non-reciprocal systems. Non-reciprocity, in this context, is given when the metastructure receives radiation from a given direction and reflects it along a second direction, but the opposite situation is not given. In other words, time reversal symmetry is broken \cite{Hadad2015}. Plethora of works have been published remarking this property \cite{Hadad2016, Sounas2017}. Other works benefits from this property to derive some applications, such as \cite{Yin2022, Chamanara2017}. Some other applications use temporal systems to tailor frequency modulation, as those in \cite{Salary2018, Shi2016, Zhang2018} for wavefront control, the one in \cite{Fang2022} for direction-of-arrival (DOE) estimation, or \cite{Wang2019} for multiplexing. 

The analysis and design of temporal and spatiotemporal models lacks from the existence of commercial tools. Generally, a home-made numerical code based on Finite-Differences Time Domain (FDTD) \cite{Itoh1989} provides good physical insight, especially to simulate the time evolution of systems. However it lacks from providing specific parameters such as reflection and transmission coefficients, and most refined techniques are commonly needed. For example, for time-varying dielectric slabs, \cite{Zurita2009} employs a classical mode matching technique whereas \cite{Sotoodehfar2022} solves the eigenvalue problem of the dispersion equation. Mode matching technique is also applied in \cite{Taravati2019} for a spacetime periodic grating. Some other techniques depart from a Floquet analysis, implementing the SD-TW modulation \cite{Tiukuvaara2021}, or modelling a Huygens metasurface with Lorentzian dispersion models \cite{Wu2020}. In \cite{Bass2022}, the solution of the system is found thanks to the implementation of a Method-of-Moments (MoM). Circuit-model interpretations have also been reported based on shunt topologies derived from admittance matrices \cite{Wang2020, Mostafa2021}. 
In most cases, direct comparisons with results provided by FDTD are needed to validate the models \cite{Taravati2022, CalozFDTD}. 


State-of-art works in the literature have traditionally focused on the study of dielectric-like systems with spatiotemporal variations. This work discusses on time-varying systems of metallic nature, consisting of infinitely extended metallic screens that appear and vanish periodically. This situation emulates systems in microwave, millimeter and low-THz regions, where metals behave as good conductors. A plane wave impinges on the structure, interacting with the time-modulated screen. Despite the simplicity of the scenario, it is prone to pose the problem in terms of Floquet and a subsequent circuit analysis. The derivation of simple expressions for the circuit elements allows the reader to acquire substantial physical insight. The methodology is motivated by previous analysis reported in the literature, as in \cite{Berral2012, Berral2012_2, Molero2017_PRE} for 1-D structures and \cite{Berral2015, Molero2017, Alex2021, Dubrovka2006} for 2-D structures modulated in space. The structure considered in this manuscript is a kind of analogous but in time. The conclusions extracted from this work can easily be extended to other higher-dimensional structures modulated in both time and spaces. 

The time-periodic system presented here is formed by a screen that periodically alternates between ``metal" and ``air" states. To recreate this system in a real-world implementation, we would require the use of materials, of metallic/semiconductor nature, that can be electronically reconfigured. 2-D materials such as graphene, molibdenum disulfide (MoS2) and hexagonal boron nitride (hBN) could be an interesting option to consider \cite{Yu2020, LI201533, Laturia2018}. For instance, it is well known that graphene can act as a metal (good conductor) when it is electrically biased. Moreover, the sheet resistance of graphene can be reconfigured depending on the bias \cite{Graphene_Growth1, Graphene_Growth2, Graphene3}. Thus, graphene can act as a good conductor (metal state), as a bad conductor, as an absorber and as a transparent layer that is perfectly matched to the surrounding media \cite{Allen2010, Zhu_2014}. Transparent-like responses would lead to the realization of the air state. Naturally, a periodic tuning of the graphene's sheet resistance to achieve metal and air states would require appropriate and specific control electronics. Alternatively, up to microwave frequencies, the use of conventional electronically-reconfigurable schemes based on PIN and varactor diodes could be considered. To recreate ``air" and ``metal" states in our time-periodic system, we would require of the realization of a reconfigurable planar frequency selective surface (FSS). When properly designed, the reconfigurable FSS would allow full transmission (air state) and full reflection (metal state) of the incident electromagnetic waves \cite{Hu2022}. In a metallic FSS, full transmission and reflection can be achieved by means of resonances in a particular range of frequencies.

The paper is organized as follows: Section II is left for the exposition of the theoretical background, analysis of the dynamic and dispersion response, and circuit derivation. Section III is used to evaluate and validate the proposal. The conclusions of the work are found at the end of the paper.

\section{Theoretical Formalism}

\begin{figure}[!t]
	\centering
		\subfigure[]{
		\includegraphics[width=0.18\textwidth]{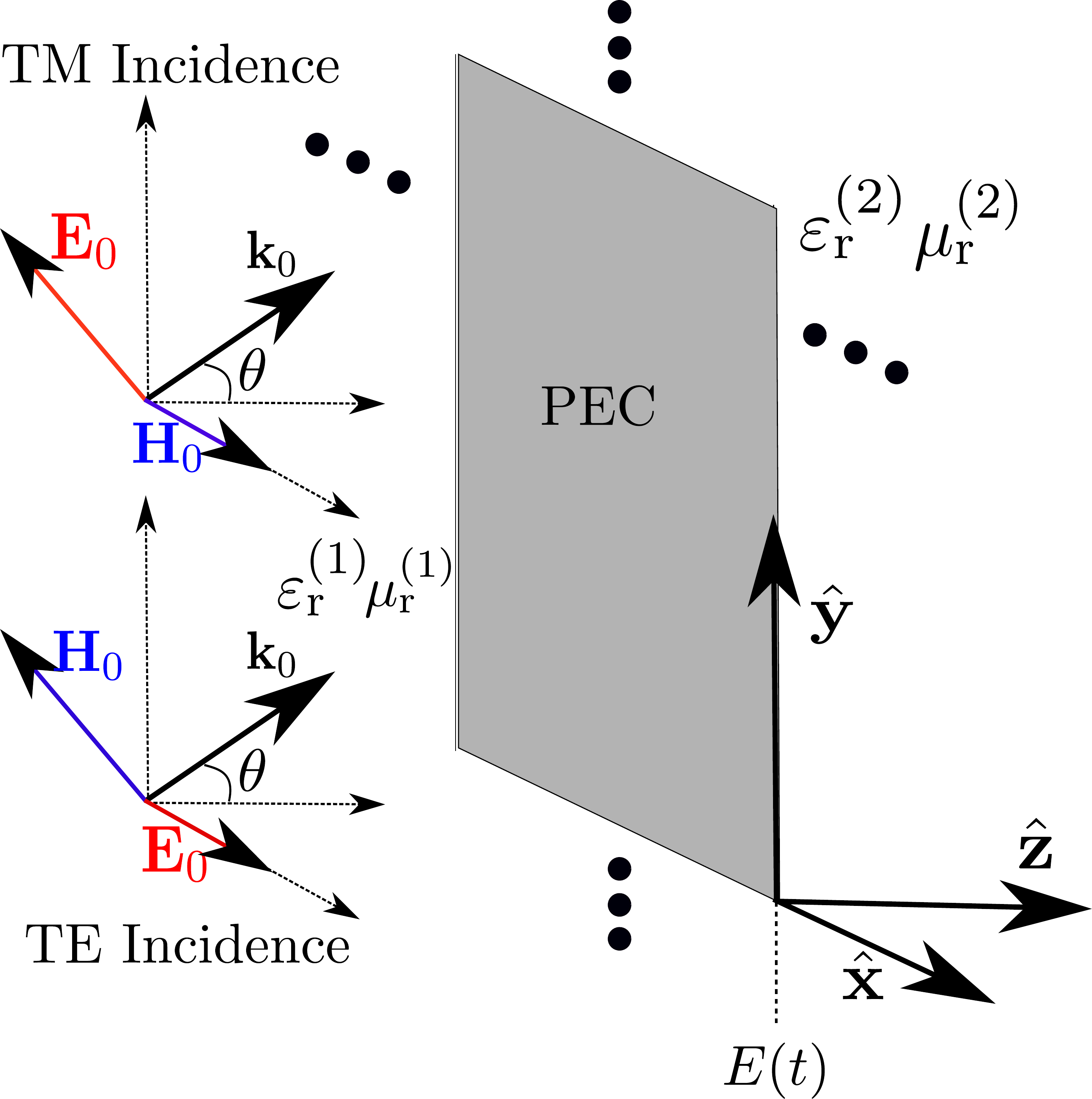}}
	\subfigure[]{
		\includegraphics[width=0.18\textwidth]{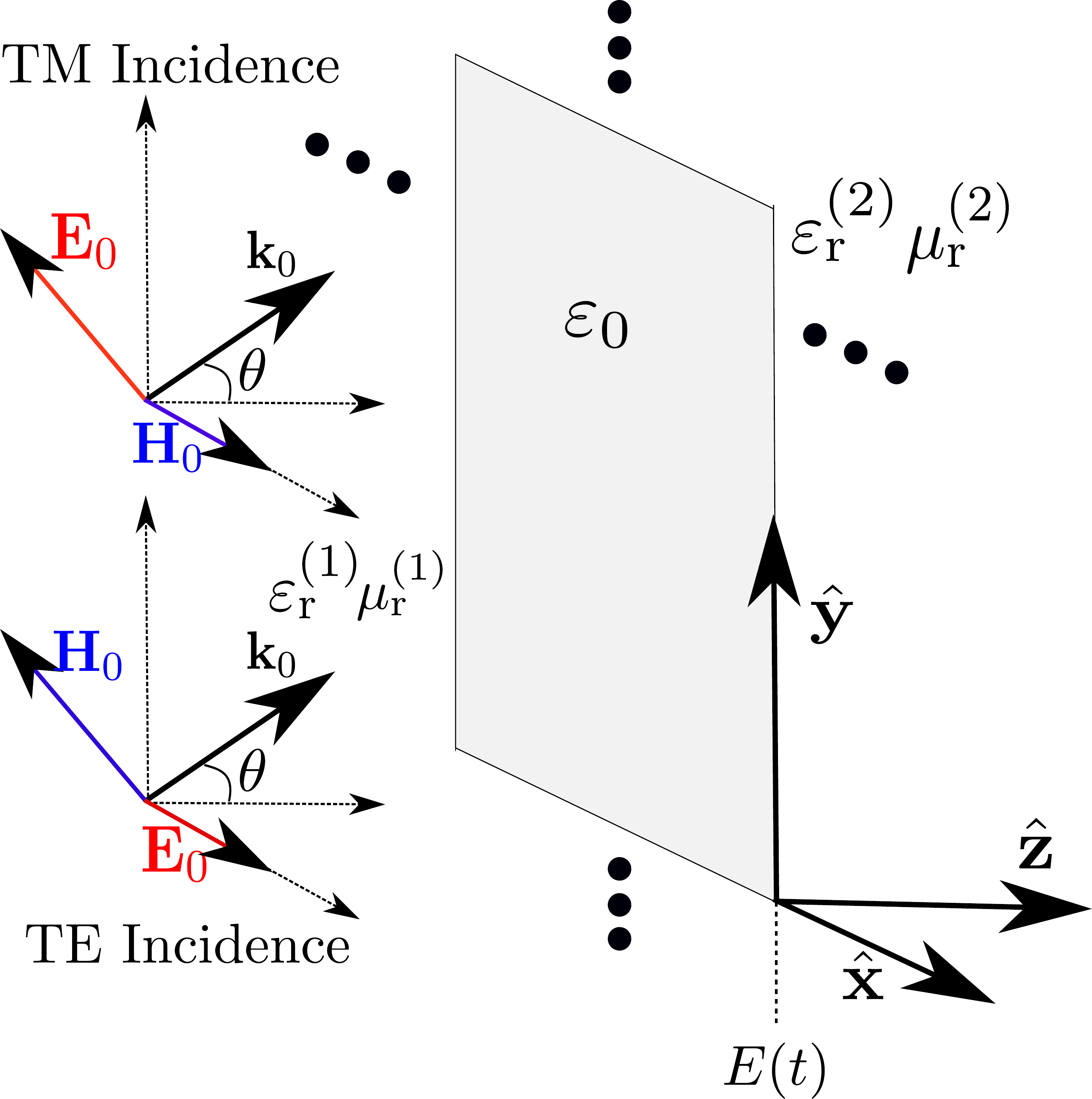}} 
	\caption{\small Sketch of the scenario. The time interface is found at position $z = 0$. (a) Time interval when the interface is a PEC sheet (``metal" state). (b) Time interval when the interface vanishes (``air" state).}
	\label{Scenario}
\end{figure}

The scenario to evaluate a time-varying metallic structure is depicted in \Fig{Scenario}. An  infinitely-extended time-varying interface is placed in between of two semi-infinite media ($i=1$ for input, $i=2$ for output). The interface (or discontinuity) is a perfect-electric conductor (PEC) sheet that appears and vanishes periodically with period $T_{\text{s}}$ and angular frequency $\omega_{\text{s}} = 2\pi / T_{\text{s}}$. A plane wave vibrating with period of $T_{0}$ and angular frequency $\omega_{0} = 2\pi/T_{0}$ illuminates the discontinuity. 
A real implementation of a system of this kind can be, for example, a graphene sheet switching between conductor and non-conductor states, controlled by an external bias voltage \cite{Graphene_Growth1, Graphene_Growth2, Graphene3}. 

Let us define an arbitrary plane wave impinging on such a discontinuity. Assuming TE incidence, the fields associated to this plane wave are expressed as
\begin{align}
\mathbf{E}_{0} &= \text{e}^{\text{j} \omega_{0} t - \text{j} k_{t} y - \text{j}\beta_{0}^{(1)} z} \, \hat{{\mathbf{x}}} \\
\mathbf{H}_{0} &= {Y_{0}^{(1)}}\, \text{e}^{\text{j} \omega_{0} t - \text{j} k_{t} y - \text{j}\beta_{0}^{(1)} z} \, \left[\hat{{\mathbf{y}}} \cos (\theta) - \hat{{\mathbf{z}}} \sin (\theta) \right] 
\end{align}
where we have assumed electric-field amplitude unity. The parameter $Y_{0}^{(1)}$ is the wave admittance in the incidence region $(1)$ ($z < 0$), and the vectors \linebreak $k_{t} = \sqrt{\varepsilon_r^{(1)}\mu_r^{(1)}} \, k_{0} \sin(\theta)$ and $\beta_{0}^{(1)} = \sqrt{\varepsilon_r^{(1)}\mu_r^{(1)}} \,  k_{0} \cos(\theta)$ refer to the transverse and longitudinal components of the incident wavevector, respectively. The angle $\theta$ corresponds to the incidence angle. In the case of TM incidence, the fields of the incident plane wave are expressed as
\begin{align}
\mathbf{E}_{0} &= \text{e}^{\text{j} \omega_{0} t - \text{j} k_{t} y - \text{j}\beta_{0}^{(1)} z} \, \left[\hat{{\mathbf{y}}} \cos (\theta) - \hat{{\mathbf{z}}} \sin (\theta) \right] \\
\mathbf{H}_{0} &= {Y_{0}^{(1)}}\,  \text{e}^{\text{j} \omega_{0} t - \text{j} k_{t} y - \text{j}\beta_{0}^{(1)} z} \, \hat{{\mathbf{x}}}
\end{align}

\subsection{Floquet-Bloch Expansion}

Due to the existence of the time-varying interface and its interaction with the incidence wave, the global transverse electromagnetic field in region $(1)$ admits to be represented in terms of a Floquet series:
\begin{multline}\label{E1}
    \mathbf{E}_{\text{t}}^{(1)}(y, z, t) =  \bigg[\text{e}^{\text{j} \omega_{0} t - \text{j} k_{t} y - \text{j}\beta_{0}^{(1)} z} + R \text{e}^{\text{j} \omega_{0} t - \text{j} k_{t} y + \text{j}\beta_{0}^{(1)} z} \\ +  \displaystyle\sum _{\forall n \ne 0} E_{n}^{(1)} \text{e}^{\text{j} \omega_{n} t - \text{j} k_{t} y + \text{j}\beta_{n}^{(1)} z} \bigg] \hat{\mathbf{x}}
\end{multline}
\begin{multline}\label{H1}
    \mathbf{H}_{\text{t}}^{(1)}(y, z, t) =  \bigg[ Y_{0}^{(1)}\text{e}^{\text{j} \omega_{0} t - \text{j} k_{t} y - \text{j}\beta_{0}^{(1)} z}  \\ - R Y_{0}^{(1)} \text{e}^{\text{j} \omega_{0} t - \text{j} k_{t} y + \text{j}\beta_{0}^{(1)} z}  -  \displaystyle\sum _{\forall n \ne 0} Y_{n}^{(1)} E_{n}^{(1)} \text{e}^{\text{j} \omega_{n} t - \text{j} k_{t} y + \text{j}\beta_{n}^{(1)} z} \bigg]\hat{\mathbf{y}}
\end{multline}
where $E_{n}^{(1)}$ is the amplitude of the $n$th harmonic, $R$ is the reflection coefficient caused by the time-varying interface, and $\omega_{n}$ is the angular frequency associated with the $n$th-order harmonic:
\begin{equation} \label{wn}
    \omega_{n} = \omega_{0} + n \omega_{\text{s}} \,.
\end{equation}
As it can be appreciated in eqs.~\eqref{E1}-\eqref{wn}, Floquet harmonics include a dependence not only on the angular rate of change $\omega_s$, related to the time-varying metallic screen, but also on the angular frequency of the incident plane wave, $\omega_0$. The additional phase factor that $\omega_0$ brings to the series provokes that the problem cannot be reduced to a conventional Fourier series.

In a similar way, we define the transverse electromagnetic field at the region $(2)$ ($z > 0$), 
\begin{multline}\label{E2}
    \mathbf{E}_{\text{t}}^{(2)}(y, z, t) = \bigg[ T \text{e}^{\text{j} \omega_{0} t - \text{j} k_{t} y - \text{j}\beta_{0}^{(2)} z} \\ +  \displaystyle\sum _{\forall n \ne 0} E_{n}^{(2)} \text{e}^{\text{j} \omega_{n} t - \text{j} k_{t} y - \text{j}\beta_{n}^{(2)} z} \bigg] \hat{\mathbf{x}}
\end{multline}
\begin{multline}\label{H2}
    \mathbf{H}_{\text{t}}^{(2)}(y, z, t) =  \bigg[ T Y_{0}^{(2)} \text{e}^{\text{j} \omega_{0} t - \text{j} k_{t} y - \text{j}\beta_{0}^{(2)} z} \\ +  \displaystyle\sum _{\forall n \ne 0} Y_{n}^{(2)} E_{n}^{(2)} \text{e}^{\text{j} \omega_{n} t - \text{j} k_{t} y - \text{j}\beta_{n}^{(2)} z} \bigg]\hat{\mathbf{y}}
\end{multline}
where $T$ denotes the transmission coefficient and $E_{n}^{(2)}$ the amplitude associated to a $n$th-order Floquet harmonic. 

In the former expressions, $Y_{n}^{(i)}$ is the admittance of the $i$-th medium ($i=1$ for input, $i=2$ for output) associated with the $n$th-order harmonic:
\begin{align}
\label{admittanceTM}Y_{n}^{(i)} &= \frac{\varepsilon_r^{(i)} \varepsilon_0\, \omega_n}{ \beta_{n}^{(i)}} \hspace{5 mm} \text{TM incidence}\\ 
\label{admittanceTE}Y_{n}^{(i)} &= \frac{\beta_{n}^{(i)}}{\mu_r^{(i)} \mu_0\, \omega_n} \hspace{5 mm} \text{TE incidence}
\end{align}
with 
\begin{align}
    \label{beta} \beta_{n}^{(i)} &= 
    \sqrt{\left[k_{n}^{(i)}\right]^2 - k_{t}^2} \\
    \label{kn} k_{n}^{(i)} &= \sqrt{\varepsilon_r^{(i)}\mu_r^{(i)}} \, \frac{\omega_{n}}{c}\,.
\end{align}
The parameter $k_{n}^{(i)}$ is the $n$th-order wavenumber and $\beta_{n}^{(i)}$ its corresponding propagation constant, both in region $(i)$. For TE incidence, just TE admittances in \eqref{admittanceTE} take part in \eqref{H1} and \eqref{H2}. Similarly, for TM incidence \eqref{H1} and \eqref{H2} do use TM admittances only.

\subsection{Dispersion Properties}
\begin{figure}[!t]
\centering
	\subfigure{\includegraphics[width=0.38\textwidth]{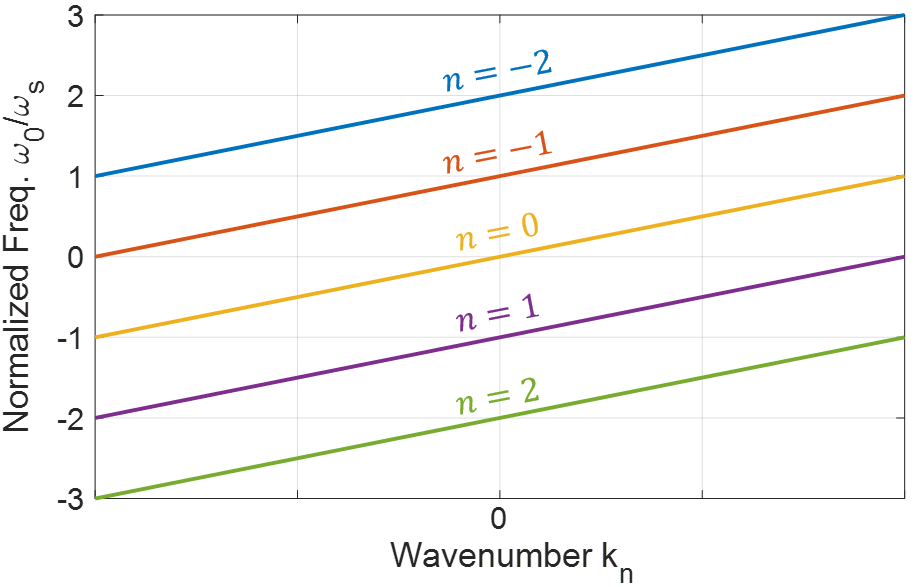}}
	\caption{\small Dispersion curves for the time-modulated metallic screen.} 
\label{dispersion}
\end{figure}

Figure \ref{dispersion} illustrates the dispersion relation $\omega(k)$ for the time-modulated metallic screen. Dispersion curves are linear (non dispersive) for all the considered harmonics $n$ and are separated from each other by $\omega_s$. Moreover, no stopband region is observed. This is due to the infinitesimal thickness of the time-modulated screen. Actually, this situation is expected to hold as long as the thickness of the screen is much smaller compared to the incident wavelength. This is a remarkable difference compared to time-modulated dielectrics slabs $\varepsilon(t)$ \cite{Zurita2009, Sotoodehfar2022}, which are typically dispersive and present forbidden bands. Another fact to remark is the existence of negative frequencies and wavenumbers. This means that the corresponding $n$th-order wavenumber is negative, referring to a mode travelling backward \cite{Taravati2019}.

Additionally, by looking at eqs. \eqref{beta}, \eqref{kn}, it can be readily inferred that most of the diffraction orders created by time-varying metallic screen are purely propagative. It can be demonstrated that only a few negative integer orders $n$ are evanescent ($\beta_n$ should be imaginary for evanescent waves). These orders would be restricted to
\begin{equation} \label{n_evanescent}
    -\frac{\omega_0}{\omega_s} \left[1 + \sin (\theta) \right] < n < -\frac{\omega_0}{\omega_s} \left[1 - \sin (\theta) \right], \, \, \textrm{Eva. waves},
\end{equation}
when the considered background is air. In fact, in case of normal incidence ($\theta = 0$), there are no evanescent waves. Conversely, larger ratios $\omega_0/\omega_s$ in combination with large incident angles $\theta$ provoke that a greater number of modes are evanescent.

\subsection{Time-varying E-field at the Discontinuity. Equivalent-circuit interpretation}

In order to proceed, we will assume the a priori  knowledge of the time-dependent field profile at the discontinuity. Without loss of generality, it can be described by a function depending on time  $\mathbf{E}(t)$. This assumption allows us to apply and adapt the models presented in \cite{Berral2012, Berral2012_2} for a time periodic problem.  

Thus, we first begin by imposing the continuity of the electric field at the interface $z = 0$,
\begin{equation}
    \mathbf{E}^{(1)}(y, 0, t) = \mathbf{E}^{(2)}(y, 0, t) = \mathbf{E}(t)\,,
\end{equation}
where both $\mathbf{E}^{(1)}(y, 0, t)$ and $\mathbf{E}^{(2)}(y, 0, t)$ have the field profile at the discontinuity $\mathbf{E}(t)$. This equality allows us to apply standard Fourier analysis, where we reach the following expressions:
\begin{align}\label{orden0}
    (1 + R) &= T = \frac{1}{T_{\text{s}}} \displaystyle\int_{-T_{\text{s}}/2} ^{T_{\text{s}}/2} E(t) \text{e}^{-\text{j} \omega_{0} t} \text{d}t \\
    \label{ordenn} E_{n}^{(1)} &= E_{n}^{(2)} = \frac{1}{T_{\text{s}}} \displaystyle\int_{-T_{\text{s}}/2} ^{T_{\text{s}}/2} E(t) \text{e}^{-\text{j} \omega_{n} t} \text{d}t\,.
\end{align}
From eq.\eqref{orden0} and eq.\eqref{ordenn} we get the following relationship:
\begin{equation}\label{relation}
    E_{n}^{(1)} = E_{n}^{(2)} = (1 + R) N(\omega_{n}) 
\end{equation}
where
\begin{equation}\label{Nn}
N(\omega_{n}) = \frac{\displaystyle\int_{-T_{\text{s}}/2} ^{T_{\text{s}}/2} E(t) \text{e}^{-\text{j} \omega_{n} t} \text{d}t}{\displaystyle\int_{-T_{\text{s}}/2} ^{T_{\text{s}}/2} E(t) \text{e}^{-\text{j} \omega_{0} t} \text{d}t}    
\end{equation}
accounts for the coupling between the incident wave and the corresponding $n$th-order harmonic. 

\begin{figure}[!t]
\centering
	\subfigure[]{\hspace{-1cm}\includegraphics[width=0.38\textwidth]{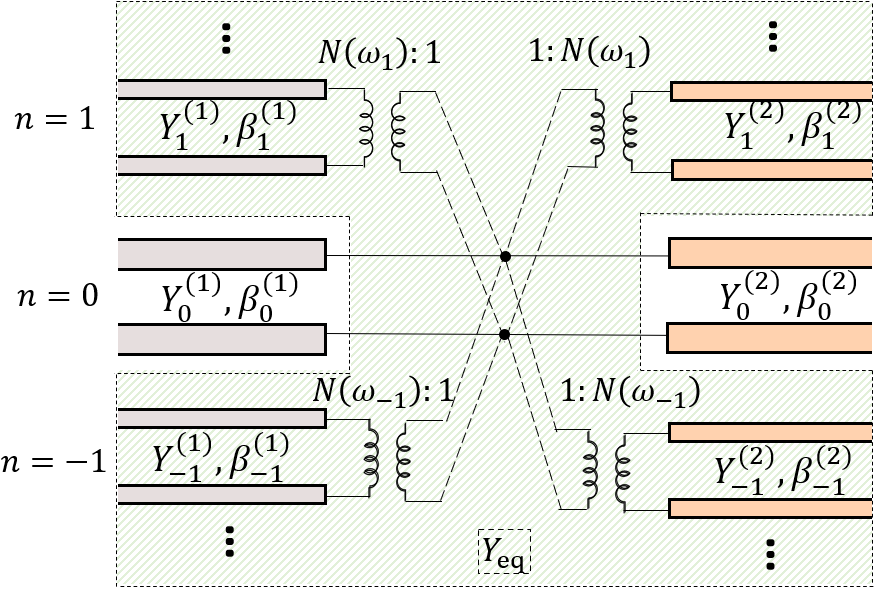}}
	\subfigure[]{\hspace{-0.9cm}\includegraphics[width=0.35\textwidth]{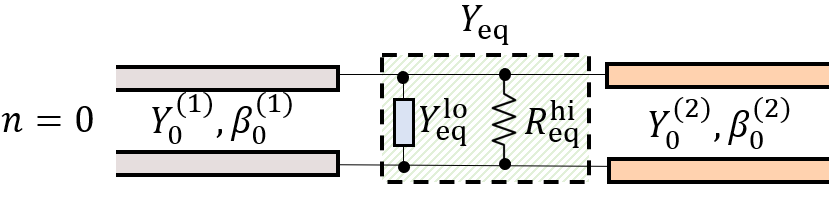}}
	\caption{\small (a) Equivalent circuit that models the time-modulated metallic screen. It consists of a infinite parallel connection of transmission lines, each representing a different harmonic. (b) Compact equivalent circuit. The equivalent admittance $Y_\mathrm{eq}$, which contains all relevant information about the time-varying screen, can be subdivided into low-order (lo) and higher-order (hi) contributions. } 
\label{circuitos}
\end{figure}

Now, the continuity of the instantaneous Poynting vector at the interface is imposed. The power passing through the interface is evaluated over a period $T_{\text{s}}$, 
\begin{equation}
\displaystyle\int_{-T_{\text{s}}/2}^{T_{\text{s}}/2} \mathbf{E}(t) \times \mathbf{H}^{(1)}(y, 0, t) \text{d}t = \displaystyle\int_{-T_{\text{s}}/2}^{T_{\text{s}}/2} \mathbf{E}(t) \times \mathbf{H}^{(2)}(y, 0 , t) \text{d}t \, 
\end{equation}
leading to
\begin{multline}\label{long}
    (1 - R)Y_{0}^{(1)} \displaystyle\int_{-T_{\text{s}}/2}^{T_{\text{s}}/2} E(t) \text{e}^{\text{j} \omega_{0}t} \text{d}t \\ - (1 + R) \displaystyle\sum_{\forall n \ne 0} N(\omega_{n}) Y_{n}^{(1)} \displaystyle\int _{-T_{\text{s}}/2}^{T_{\text{s}}/2} E(t) \text{e}^{\text{j} \omega_{n} t} \text{d}t \\ = (1 + R)Y_{0}^{(2)} \displaystyle\int_{-T_{\text{s}}/2}^{T_{\text{s}}/2} E(t) \text{e}^{\text{j} \omega_{0}t} \text{d}t \\ + (1 + R) \displaystyle\sum_{\forall n \ne 0} N(\omega_{n}) Y_{n}^{(2)} \displaystyle\int _{-T_{\text{s}}/2}^{T_{\text{s}}/2} E(t) \text{e}^{\text{j} \omega_{n} t} \text{d}t\,.
\end{multline}
{The former expression is valid as long as the input and output media are identical}. Comparing the integrals in eq.\eqref{long} and rearranging terms, the reflection coefficient is finally expressed in the following way,
\begin{equation}\label{R}
    R = \frac{Y_{0}^{(1)} - Y_{0}^{(2)} - \displaystyle\sum_{\forall n \ne 0} |N(\omega_{n})|^{2} (Y_{n}^{(1)} + Y_{n}^{(2)})}{Y_{0}^{(1)} + Y_{0}^{(2)} + \displaystyle\sum_{\forall n \ne 0} |N(\omega_{n})|^{2} (Y_{n}^{(1)} + Y_{n}^{(2)})} \,,
\end{equation}
where we can identify each of the admittances taking part on the expression as individual transmission lines with characteristic admittance $Y_{n}^{(i)}$ and propagation constant $\beta_{n}^{(i)}$. Similarly as in \cite{Berral2012, Alex2021}, eq.\eqref{R} is circuitally interpreted by the topology shown in \Fig{circuitos}. Each of the parameters $N(\omega_{n})$ is interpreted as a complex transformer (it transforms both amplitude and phase). Moreover, the infinite sum in \eqref{R} groups all relevant information about the diffracted waves created by the time-modulated screen. From a circuit standpoint, this term can be interpreted as an equivalent admittance $Y_\mathrm{eq}$ that reads
\begin{equation}
    Y_\mathrm{eq} = \displaystyle\sum_{\forall n \ne 0} |N(\omega_{n})|^{2} (Y_{n}^{(1)} + Y_{n}^{(2)})
\end{equation}
Thus, the reflection coefficient can be rewritten as
\begin{equation}\label{R2}
    R = \frac{Y_{0}^{(1)} - Y_{0}^{(2)} -  Y_\mathrm{eq}}{Y_{0}^{(1)} + Y_{0}^{(2)} +  Y_\mathrm{eq}}
\end{equation}

Close inspection of eqs. \eqref{beta} and \eqref{kn} reveals that higher-order harmonics are \emph{propagative} in the present time-modulated screen (see Appendix A for further details). This can be appreciated by looking at the wave admittances of the $n$-th higher-order mode,
\begin{align}\label{TEeqTM}
Y_{n}^{(i),\textrm{TM}} = Y_{n}^{(i),\textrm{TE}} \approx \sqrt{\frac{\varepsilon_r^{(i)} \varepsilon_0}{\mu_r^{(i)} \mu_0}} = Y_0^{(i)}, \quad |n|\gg 1\, \,,
\end{align}
which are real-valued (in lossless media), independent from index $n$, and identical for both TM and TE polarizations. This provokes that higher-order modes contribute with a purely resistive term $R_\mathrm{eq}^\textrm{hi}$  that can be modeled as  a resistor [see Figure \ref{circuitos}(b)]
\begin{equation}\label{Rho}
    \frac{1}{R_\mathrm{eq}^\textrm{hi}} = (Y_{0}^{(1)} + Y_{0}^{(2)}) \displaystyle \sum_{|n| > N_\mathrm{lo}}^{\infty} |N(\omega_{n})|^{2} 
\end{equation}
The minimum $n$th order associated with a mode participating in \eqref{Rho}, $N_\mathrm{lo}$ would be calculated by just considering \linebreak $|\beta_{n}| \gg k_{\text{t}}$. Thus the summation in \eqref{Rho} can be calculated once and stored, since it will keep invariant for any incidence angle. The former discussion shows a conceptual change with respect to spatially-modulated gratings, where higher-order harmonics are normally evanescent and carry reactive power (capacitive and inductive for TM and TE modes, respectively) \cite{Berral2012, Berral2012_2, Alex2021}.

The situation is different when considering low-order harmonics, whose associated wave impedance/admittance is a function, among other parameters, of the incident angle $\theta$. Thus, the \emph{complex} low-order contribution of the equivalent admittance present in Figure \ref{circuitos}(b) can be computed as
\begin{equation}\label{Ylo}    Y_\mathrm{eq}^\mathrm{lo} = \displaystyle\sum_{\substack{n=-N_\mathrm{lo} \\ n \neq 0}}^{n=N_\mathrm{lo}} |N(\omega_{n})|^{2} (Y_{n}^{(1)} + Y_{n}^{(2)})\,.
\end{equation}
At this point, it is important to remark that a n-$th$ TE and TM low-order harmonic do now differ. Unlike higher-order harmonics, which behaves almost identically regardless of their TM or TE nature (see \eqref{TEeqTM} for instance), admittances associated with lower-order harmonics are governed by the expressions in \eqref{admittanceTM} and \eqref{admittanceTE}, which exhibit clear differences between them. The global lower-order admittance $Y_\mathrm{eq}^\mathrm{lo}$ in \eqref{Ylo} varies according to the TE- or TM-incidence scenario, influencing both the reflection coefficient $R$ and the amplitude associated with the lower-order harmonics in both semi-spaces $E_{n}^{(1/2)}$. Thus, the resulting electromagnetic response of the whole system therefore differs with a TE- or a TM-case in a general oblique-incidence scenario. However, the behavior of the time-varying system is indistinguishable under TE or TM excitation when the incidence is normal to the discontinuity plane ($\hat{\mathbf{z}}$-direction according to the frame of coordinates in \Fig{Scenario}). TE incidence will just excite TE higher-order harmonics and TM incidence will only excite TM higher-order harmonics. This is typical in 1D problems when the transverse components of the incident wave coincide with $\hat{\mathbf{x}}$ and $\hat{\mathbf{y}}$. 

\subsection{Diffraction Angles}

The reflection/transmission angle of each $n$th-order harmonic is described by
\begin{equation}\label{angle}
    \theta_n^{(i)} = \arctan\left(\dfrac{k_{t}}{\sqrt{\varepsilon_r^{(i)} \mu_r^{(i)}[\frac{\omega_0 + n\omega_s}{c}]^2 - k_{t}^2}}\right)\,.
\end{equation}
Notice that the fundamental harmonic ($n=0$) is not affected by the time modulation. This implies that the fundamental harmonic simply obeys conventional Snell's law of refraction,
\begin{equation} \label{angle_0}
    \frac{\sin(\theta)}{\sin \left( \theta_0^{(i)} \right)} = \frac{\sqrt{\varepsilon_r^{(i)} \mu_r^{(i)}}}{\sqrt{\varepsilon_r^{(1)} \mu_r^{(1)}}}\, ,
\end{equation}
thus propagates in the same direction of incidence when the considered input and output media are air ($\theta_0^\mathrm{(i)} = \theta$).

It can be inferred from eq.\eqref{angle} that, for the case of normal incidence $\theta_{n}^{(i)} = 0, \, \forall n$. Moreover, the diffraction angle is expected to change inside the dielectric media as $\beta_n^{(i)}$ does. By looking at the expression for $\theta_n^{(i)}$, it can be appreciated that the denser the medium under consideration is, the closer the diffracted angles are to the normal (when the input media is considered to be air). This would be similar to a conventional refraction between two media. This phenomenon is even more accentuated if the time-modulated screen commutes fast between its two states; namely, if $\omega_s$ is large, the diffraction angle for higher-order harmonics goes progressively to zero as $|n|$ increases. 

\begin{figure}[!t]
	\centering
	\subfigure[]{\hspace{-0.5cm}
		\includegraphics[width=0.28\textwidth]{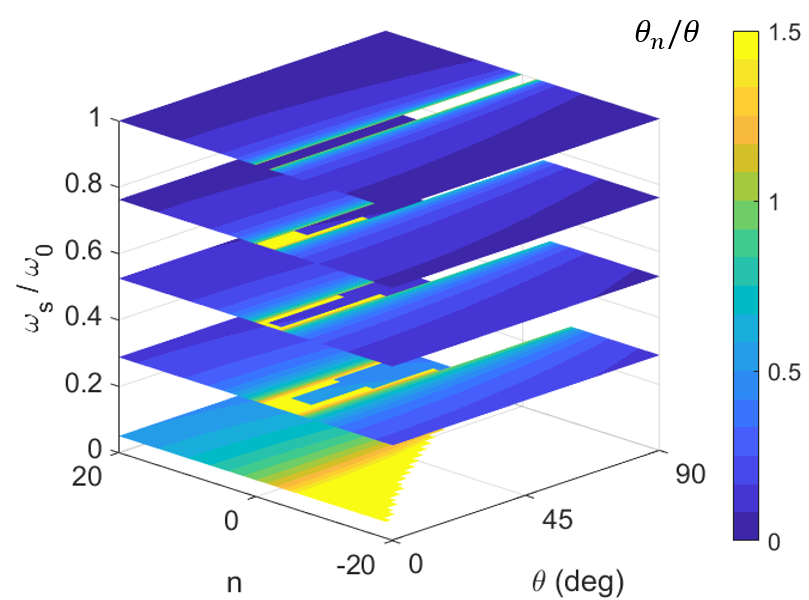}} 
	\subfigure[\hspace{-0.3cm}]{ \hspace{0.2cm}
		\includegraphics[width=0.38\linewidth, height = 0.37 \linewidth]{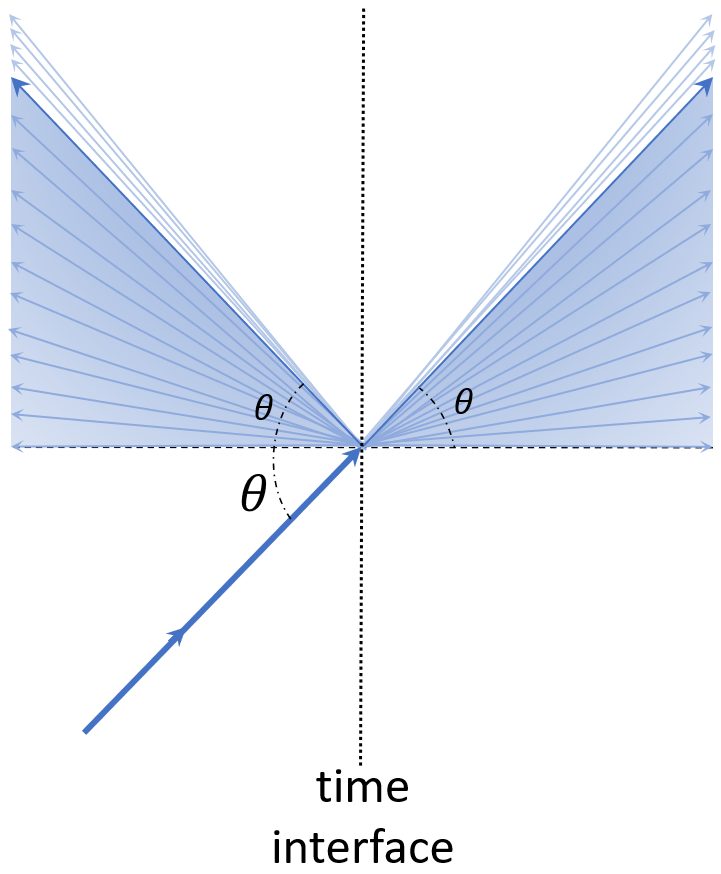}}
 	\caption{\small (a) Normalized diffraction angle ($\theta_n / \theta$) as a function of the incident angle $\theta$, harmonic index $n$, and normalized modulation frequency $\omega_s / \omega_0$. Only propagative angles are plotted. Evanescent regions are identified by blank spaces in the figure. (b) Sketch of the diffraction created by the time-modulated interface. } 
	\label{angle4D}
\end{figure}

The application of eq.\eqref{angle} has important implications from an engineering perspective. Figure \ref{angle4D}(a) illustrates a 4D representation of the (normalized) diffraction angle $\theta_n$, assuming that the time-varying screen is surrounded by air. This angle is evaluated as a function of the incident angle $\theta$, integer index $n$, and normalized modulation frequency $\omega_s / \omega_0$; namely, $\theta_n = \theta_n(\theta, n, \omega_s / \omega_0)$.  Only propagative difraction orders are considered in the figure, while evanescent orders are displayed as blank spaces. Note that in the vast majority of scenarios, the normalized diffraction angle is less than the unity, thus $\theta_n$ is generally less than $\theta$. Only in cases where $\omega_s \ll \omega_0$, the diffraction angle $\theta_n$ is larger than the incident angle $\theta$. Nonetheless, note that if $\omega_s \ll \omega_0$ in eq.\eqref{angle}, $\theta_n$ would be larger than $\theta$ but certainly close to it. On the other hand, the minimum angle at which propagative waves can diffract is $\theta_n = 0$. This is the case for higher-order harmonics ($|n \gg 1|$). Therefore, the range of diffraction angles is limited, in practice, to $\theta_n \approx [0, \theta]$. This situation is conceptually sketched in Figure \ref{angle4D}(b) for an obliquely incident plane wave that impinges the time-varying interface.

\subsection{Basis Function Choice}\label{Efunction}

\begin{figure}[!t]
	\centering
		\subfigure[]{
        \includegraphics[width=0.22\textwidth]{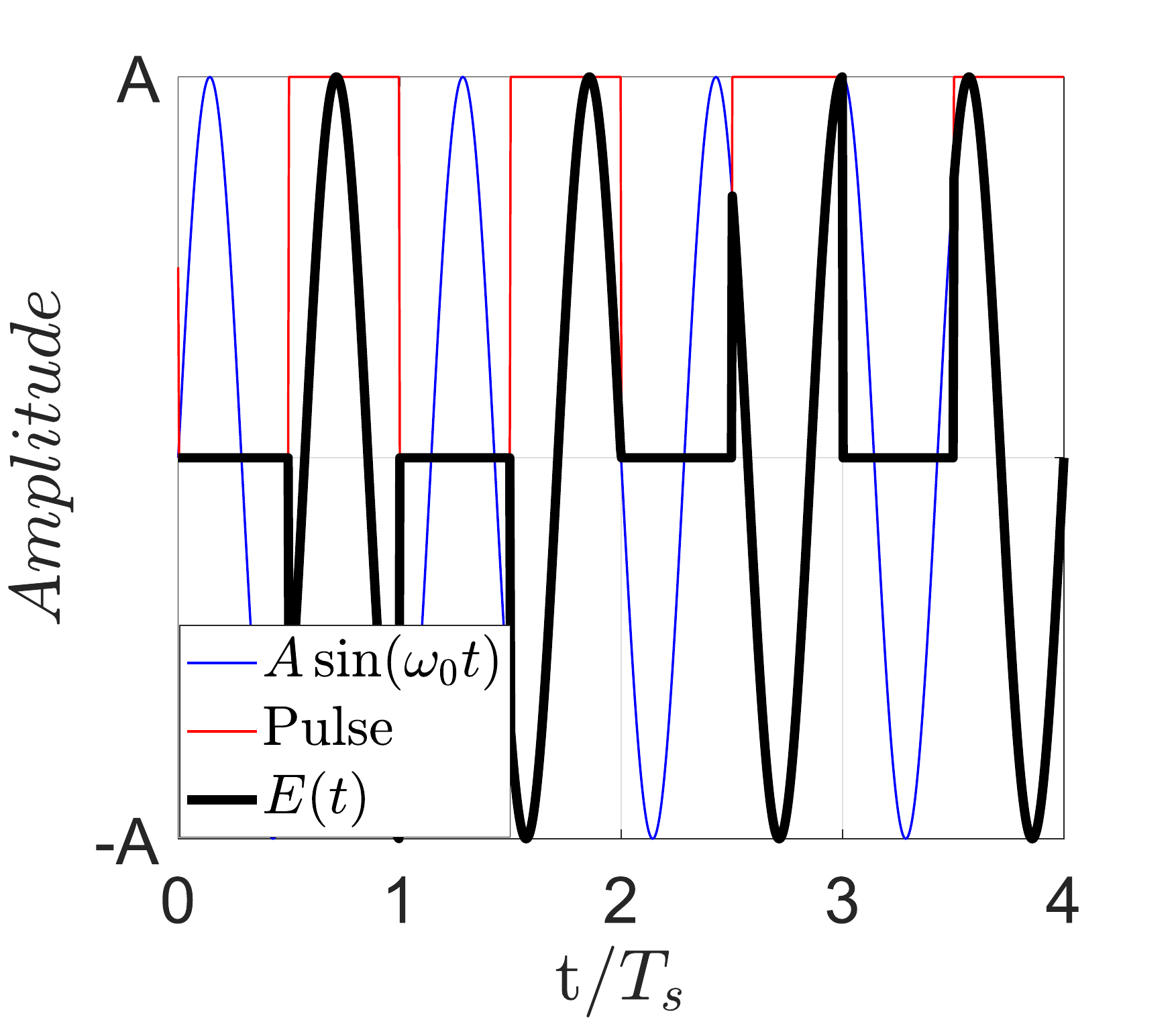}} 
	\hspace{0.2cm}	
	\subfigure[]{
        \includegraphics[width=0.22\textwidth]{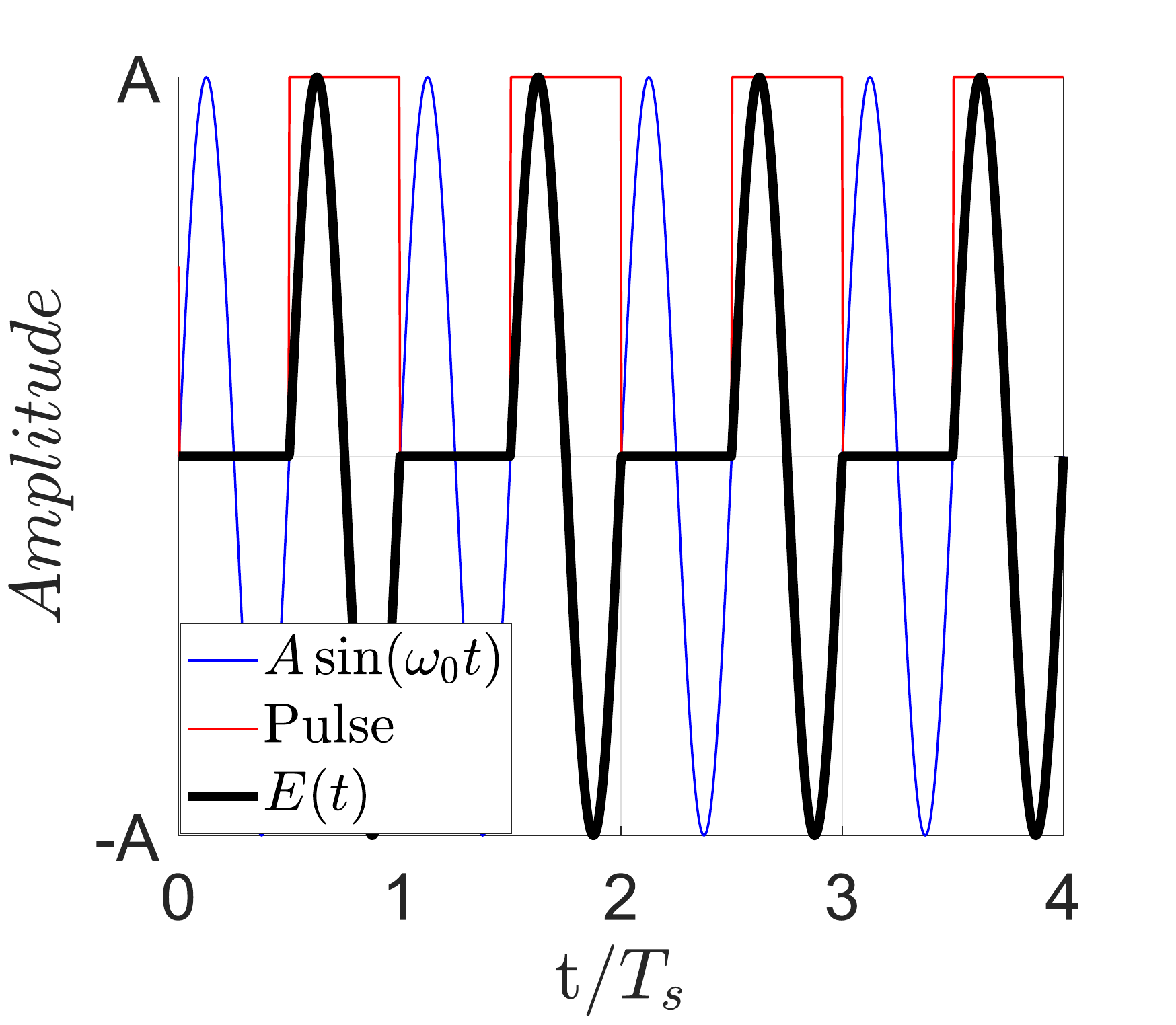}} 
	\caption{\small Field profile $E(t)$ in cases where the time modulation $\omega_s$ is slower than the vibration of the incident wave $\omega_0$. (a) $w_0/w_s = 1.75$ and (b) $w_0/w_s = 2$. } 
	\label{time_variation}
\end{figure}

The choice of the basis function $E(t)$ is crucial to validate the above approach. At this point, it is important to remark that the time period related to appearance/disappearance of the electric wall $T_{\text{s}}$ does not necessarily coincide with the period associated with $E(t)$. In order to make both periods to coincide, there must be a particular relationship between $\omega_{s}$ and $\omega_{0}$. It can be demonstrated that this relationship must satisfy $\omega_{0}/\omega_{\text{s}} = p$, with $p \in \mathbb{N}$. \Fig{time_variation} illustrates this statement in a visual way. \Fig{time_variation}(a) represents $E(t)$ for $\omega_{0}/\omega_{\text{s}} = 1.75$. We observe that $E(t)$ does not repeat within intervals separated by $T_{\text{s}}$, but by $4T_{\text{s}}$. This situation is not given when $\omega_{0}/\omega_{\text{s}} = 2$, where the periodicity is exactly $T_{\text{s}}$ as shown in \Fig{time_variation}(b). This last is the situation taken into account in this paper. It is worth remarking that the situation in \Fig{time_variation}(a) could also be addressed by considering some additional aspects, but is out of the scope of this work. 

The basis function will therefore be expressed as follows:
\begin{equation}
\mathbf{E}(t) =  \hat{\mathbf{y}} \times \left\{
0  \hspace{17 mm} -T_{\text{s}}/2 \le t < 0 \atop
A \sin(\omega_{0} t) \hspace{5 mm} 0 \le t < T_{\text{s}}/2
\right. \,.
\end{equation}
When the time-varying screen turns into a \textit{metal} \linebreak ($-T_{\text{s}}/2 \leq t < 0$),  the tangential electric field should vanish if a perfect electric conductor (PEC) is assumed. When the time-varying screen is in \textit{air state} \linebreak ($0\leq t \leq T_s/2$), the tangential field has the shape of the incident time-harmonic excitation; namely, a sinusoidal wave. Furthermore, the continuity of the electric/magnetic displacement fields $\mathbf{D}$ and $\mathbf{B}$ is guaranteed at $t = 0\,$s, what is strictly mandatory to satisfy the boundary conditions \cite{Caloz2020_1}.

Computing the Fourier transform to the basis function, we reach the following expression for the transformers \eqref{Nn},
\begin{multline} \label{Nn_expression}
N(\omega_{n}) = -\frac{4 \omega_{0}}{\omega_{n}^{2} - \omega_{0}^{2}} \times \\ \frac{ \text{e}^{-\text{j}  \omega_{n}  T_{\text{s}}/2}
\Big[\text{j} \omega_{n} \sin(\omega_{0}T_{\text{s}}/2) + \omega_{0} \cos(\omega_{0}T_{\text{s}}/2) \Big] - \omega_{0}}{\text{j} \omega_{0} T_{\text{s}} + \text{e}^{-\text{j} \omega_{0} T_{\text{s}}} - 1} \, .  
\end{multline}
%
The expressions for the transformers $N(w_n)$ will give the relative weight of the harmonics [eq. \eqref{relation}]. Note that eq. \eqref{Nn_expression} indicates that the transformers, and so the Floquet harmonics, have a decay $N(w_n) \sim 1/n $.  This weight is one of the parameters that can be compared to those extracted by FDTD.

Under the above assumption ($\omega_{0}/\omega_{\text{s}} = p$, with $p \in \mathbb{N}$), it can be appreciated that the reflection and transmission ($T = 1 + R$) coefficients do not depend on time. Both coefficients are the result of the average obtained over the period $T_s$ [see eqs. \eqref{orden0}-\eqref{ordenn}]. In general, this situation is expected to hold as long as the ratio $\omega_0 / \omega_s$ is a rational number. Nonetheless, it should be stated that rational ratios $\omega_0 / \omega_s$ would contain $M$ cycles to form a macroperiod $T = M T_\mathrm{s}$, unlike purely integer ratios $\omega_0 / \omega_s$ whose periodic response is self-contained in a single period $T_\mathrm{s}$.

\begin{figure}[!t]
	\centering
		\subfigure[]{

        \includegraphics[width=0.22\textwidth]{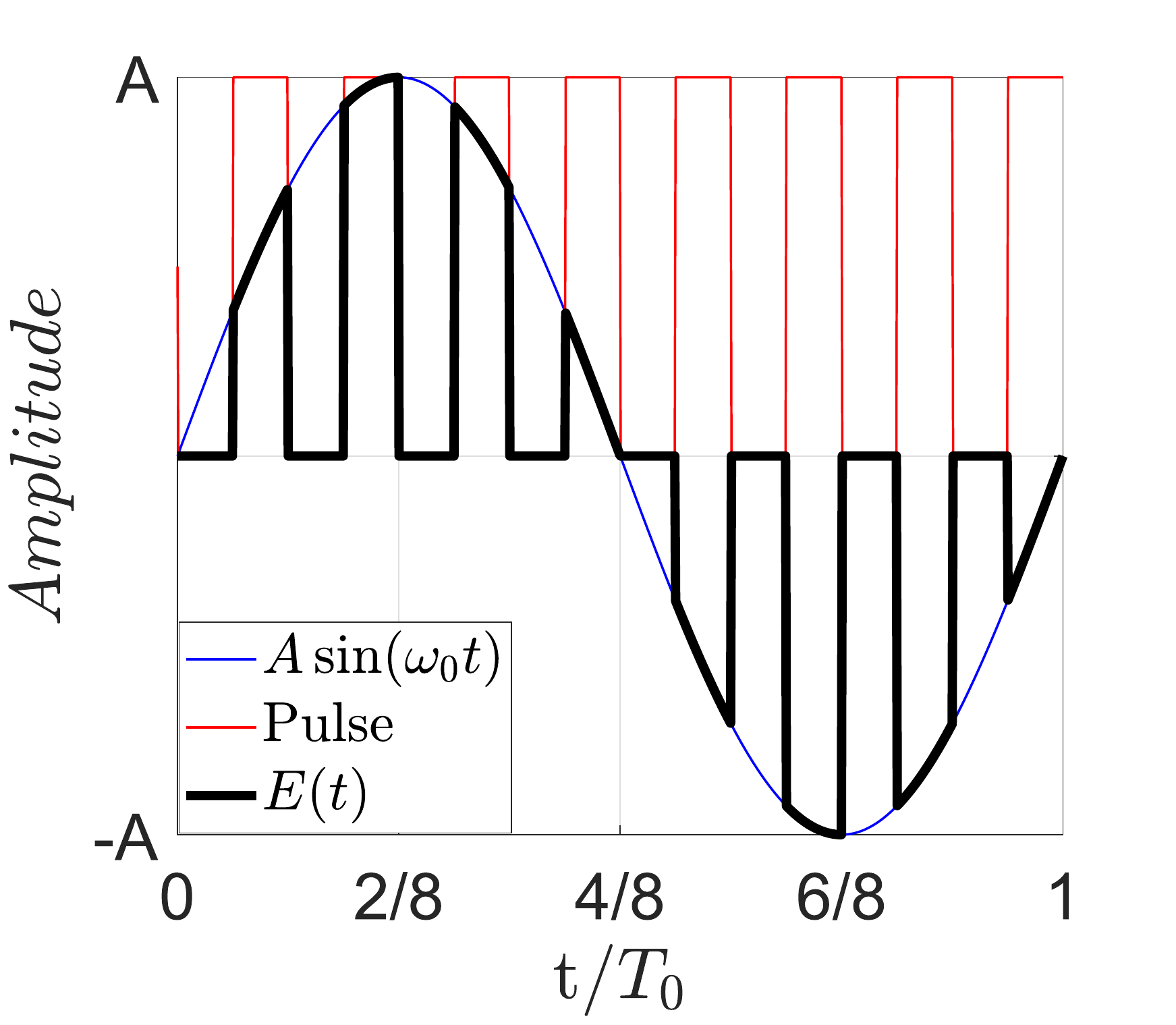}\label{f_0coma125}} 
	\hspace{0.2cm}	
	\subfigure[]{
        \includegraphics[width=0.22\textwidth]{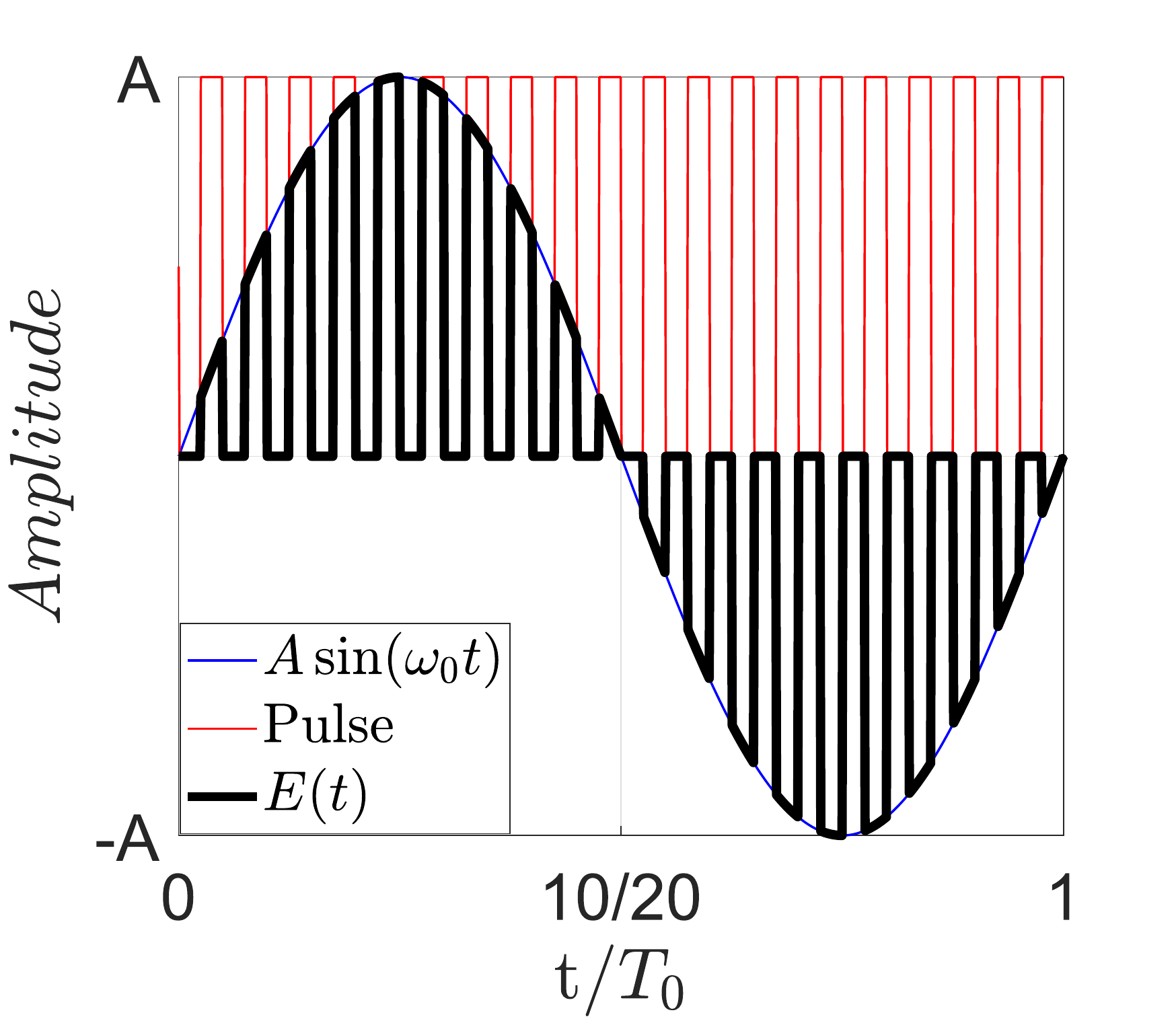}\label{f_0coma05}} 
	\caption{\small Field profile $E(t)$ in cases where  the time modulation $\omega_s$ is faster than the vibration of the incident wave $\omega_0$. (a) $w_0/w_s = 0.125$ and (b) $w_0/w_s = 0.05$. } 
	\label{time_variation_fast}
\end{figure}

Up to this point, we have only considered scenarios where the time modulation is either slower or identical to the vibration of the incident wave ($\omega_s \leq \omega_0$). Now, \Fig{time_variation_fast} presents some cases where the time modulation is faster than the vibration of the incident wave; namely, $\omega_s > \omega_0$. This scenario is a bit different from previous ones. It can be demonstrated that the time needed to achieve a complete period is now $T_0$ instead of $T_s$, as visualized in Figures \ref{time_variation_fast}(a) and \ref{time_variation_fast}(b). This is due to the fact that the period of the incident wave ($T_0$) is larger than the period of the modulation ($T_s$) in cases where $\omega_s > \omega_0$. Therefore, a sampling-like phenomena arises for the basis function $E(t)$ (black curve), caused by the rapid variation of the screen.

\subsection{Dielectric and Magnetic Losses}
Losses in dielectrics can be accounted in a straightforward manner. Simply replace the real-valued dielectric constant $\varepsilon_r^{(i)}$ by the complex term \cite{Molero2021}
\begin{equation}
    \varepsilon_r^{(i)} \rightarrow \varepsilon_r^{(i)} (1 - \mathrm{j}\tan \delta ^\mathrm{(i)})
\end{equation}
where $\tan \delta ^\mathrm{(i)}$ is the loss tangent term. A similar rationale can be applied for magnetic losses, where the relative permeability would also be defined by a complex-valued expression. Dielectric and magnetic losses can be incorporated in both models.

\section{Results, Validation \& Applications}

\begin{figure*}[!t]
	\centering
		\subfigure[]{
		\includegraphics[width=0.4\textwidth]{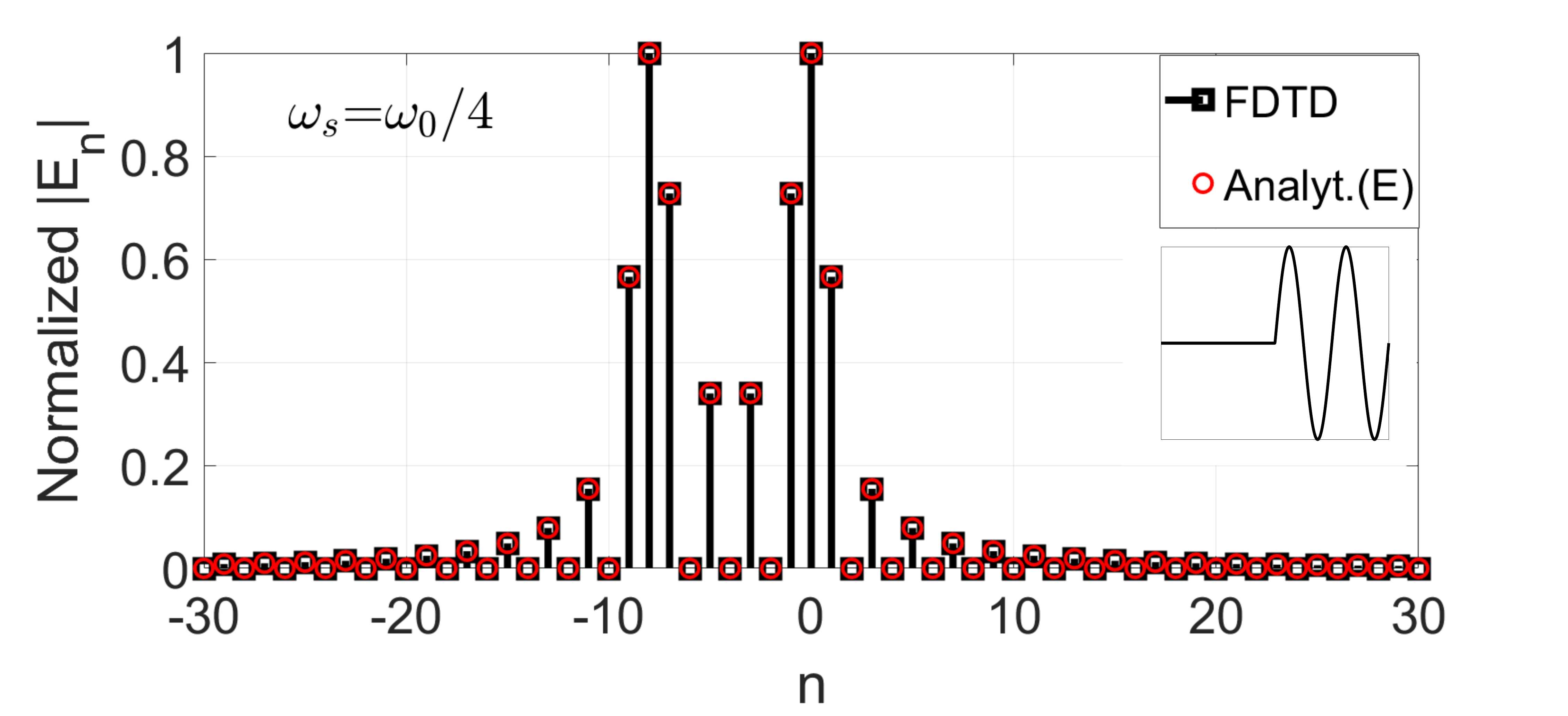}}
	\hspace{-0.5cm}	
	\subfigure[]{
		\includegraphics[width=0.4\textwidth]{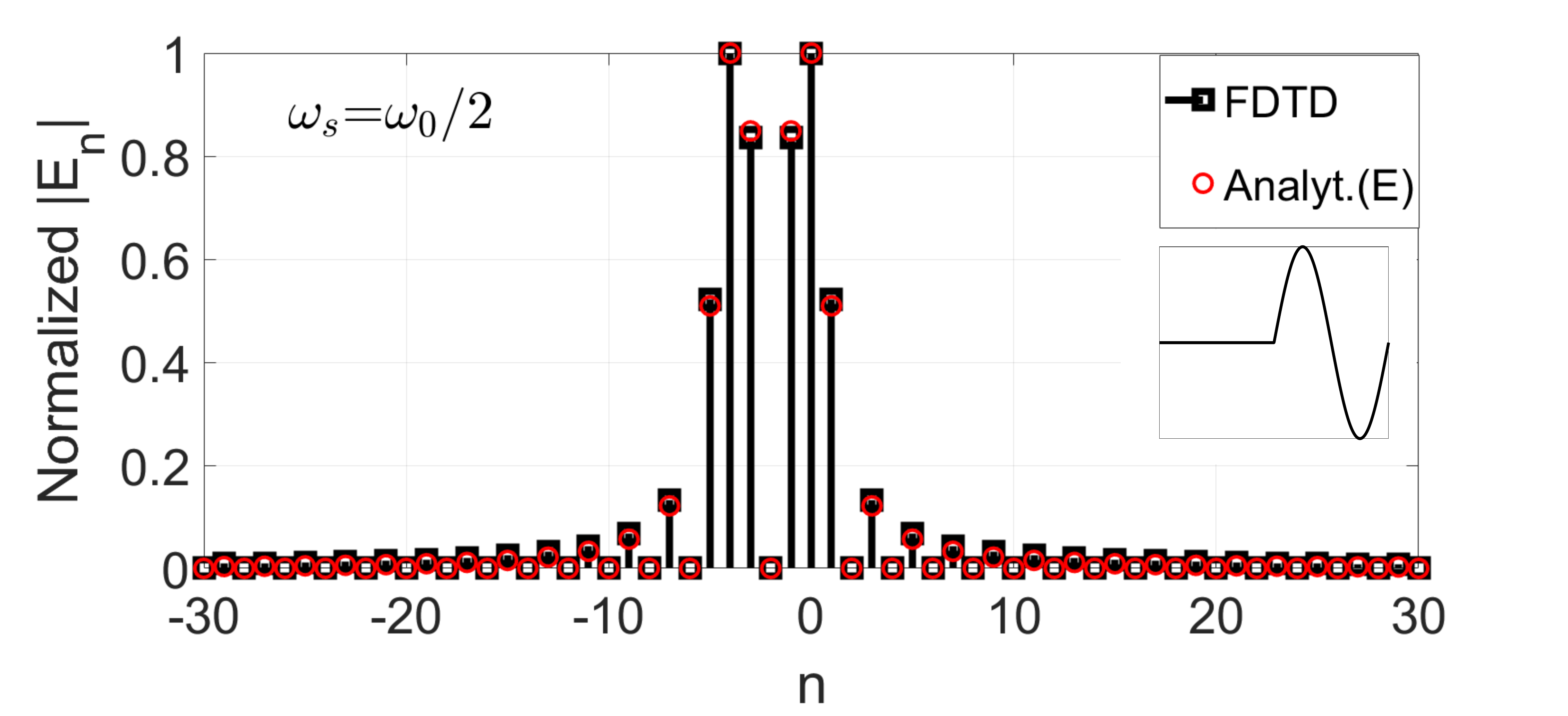}} 
	\hspace{-0.5cm}	
	\subfigure[]{
		\includegraphics[width=0.4\textwidth]{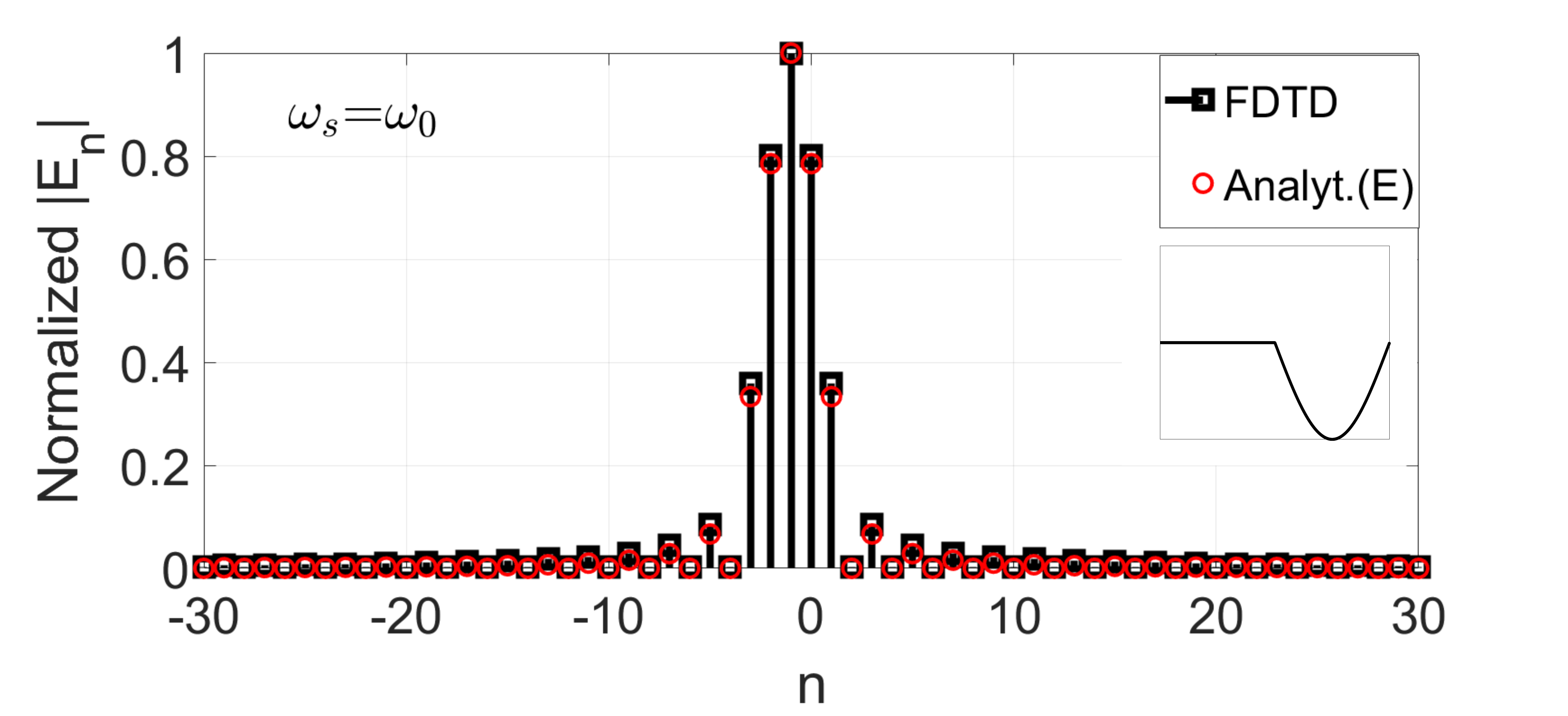}}
  	\hspace{-0.5cm}	
	\subfigure[]{
		\includegraphics[width=0.4\textwidth]{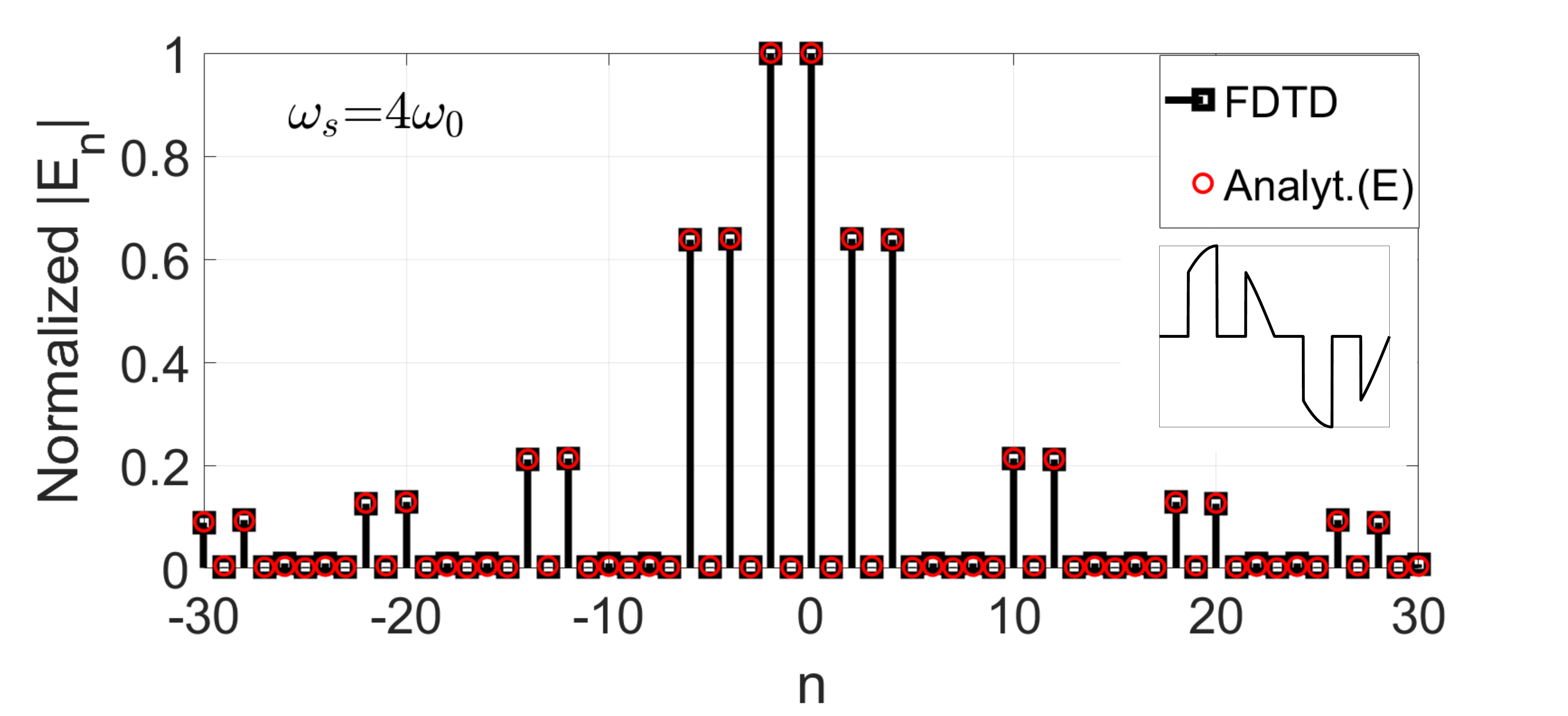}\label{N_f_0coma25_final}}
	\caption{\small Normalized Floquet coefficients $|E_n|$  when a TE-polarized plane wave impinges the time-modulated interface. Normal incidence is assumed. (a) $\omega_{\text{s}} = \omega_0/4$, (b) $\omega_{\text{s}} = \omega_0/2$, (c) $\omega_{\text{s}} = \omega_0$, (d) $\omega_{\text{s}} = 4\omega_0$. Analytical results are compared to the FDTD method.} 
	\label{Nomalcoeficients}
\end{figure*}

\begin{figure*}[!t]
	\centering
		\subfigure[]{
		\includegraphics[width=0.22\linewidth, height = 0.25 \linewidth]{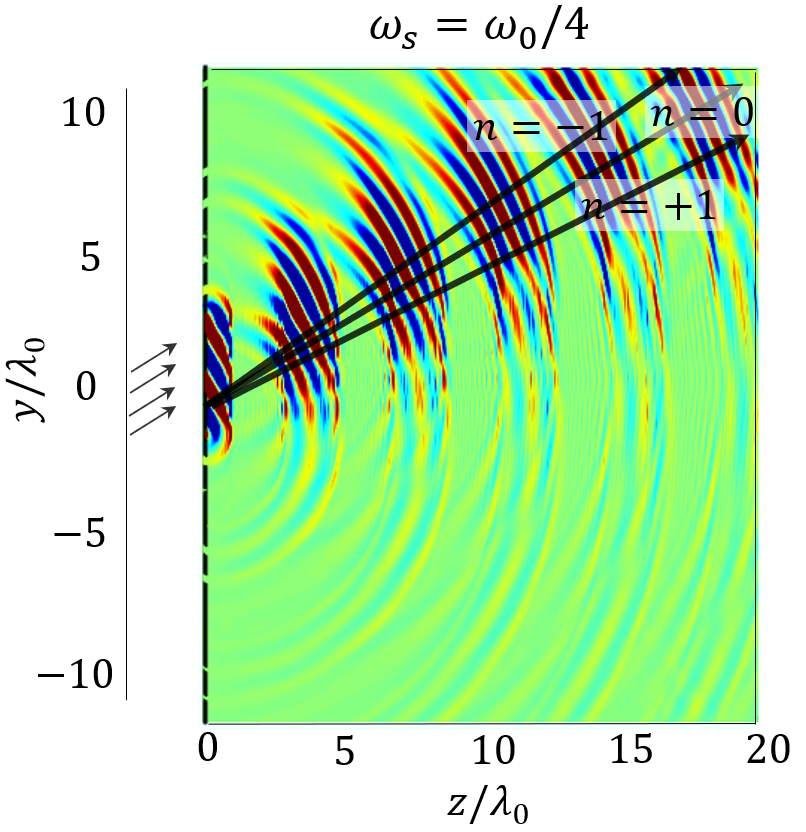}}
	\hspace{-0.2cm}	
	\subfigure[]{
		\includegraphics[width=0.225\textwidth, height = 0.25 \linewidth]{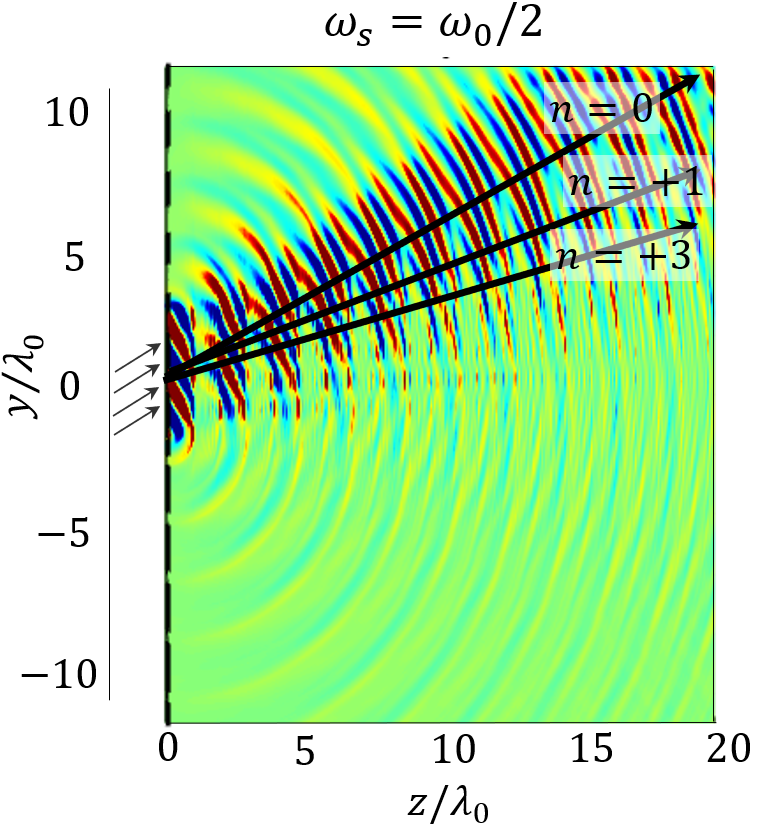}} 
	\hspace{-0.2cm}	
	\subfigure[]{
		\includegraphics[width=0.225\textwidth, height = 0.248 \linewidth]{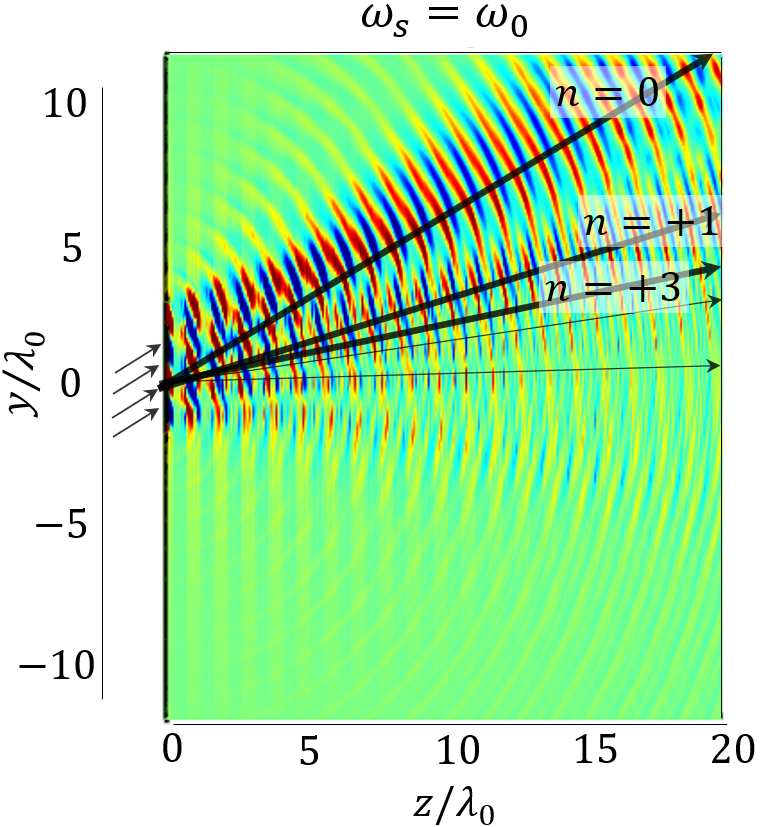}}
	\subfigure[]{
		\includegraphics[width=0.225\textwidth, height = 0.252 \linewidth]{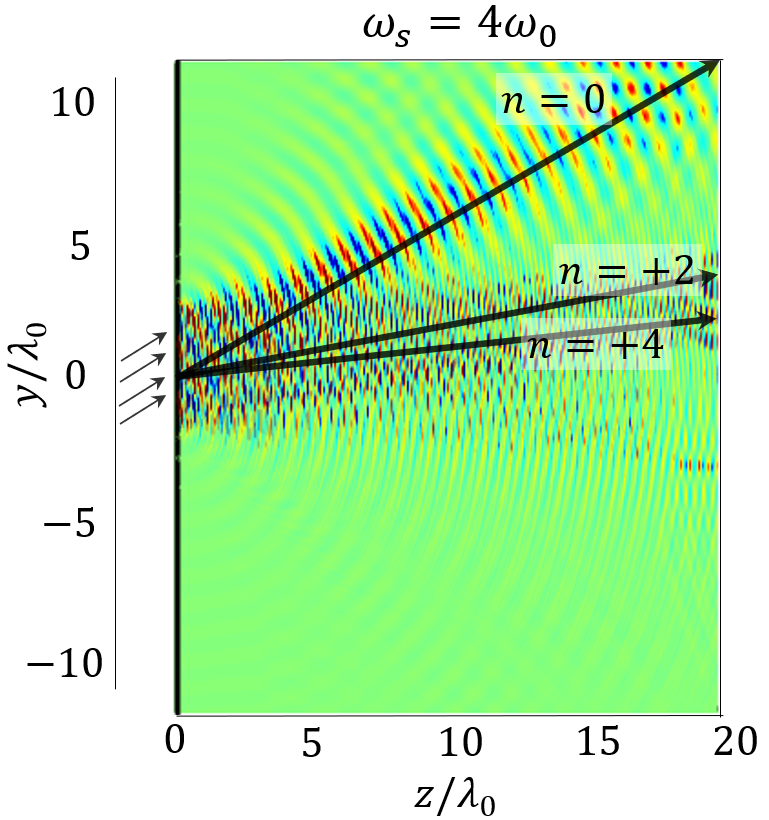}\label{oblique30_Figual4Fs}}	
	\caption{\small {FDTD simulation (electric field) showing transmission through a time-modulated metallic screen  when a TE-polarized oblique wave ($\theta = 30^\mathrm{o}$) impinges the structure. (a) $\omega_{\text{s}} = \omega_0/4$, (b) $\omega_{\text{s}} = \omega_0/2$, (c) $\omega_{\text{s}} = \omega_0$, (d) $\omega_{\text{s}} = 4\omega_0$.} }
	\label{ObliqueFDTD}
\end{figure*}

In order to validate the former approach, some results are presented here. We initially consider the scenario depicted in Figure \ref{Scenario}: a time-modulated metallic screen that periodically vanishes (PEC: $-T_s/2\leq t \leq 0$; air: $0\leq t \leq T_s/2$). Initially, both input and output media are considered to be air ($\varepsilon_r^{(1)} = \mu_r^{(1)} = \varepsilon_r^{(2)} = \mu_r^{(2)} = 1$). Nonetheless, we should remark that the former approach can be further extended to more complex scenarios involving dielectrics and modulations in both space and time. 

Figure \ref{Nomalcoeficients} illustrates the normalized Floquet coefficients $|E_n|$ extracted with the present approach and a self-implemented FDTD  formulation (see Appendix B for specific details related to the FDTD) for the case of a time-modulated screen with different modulation angular frequencies $\omega_{\text{s}}$. Normal TE incidence is assumed. As an indication, the selected basis function $E(t)$ is also included as an inset in Figures \ref{Nomalcoeficients}(a)-(c).
An excellent agreement is observed between the results extracted from the Floquet-Bloch approach and the FDTD. 

Some conclusions can be extracted by looking at \linebreak Figure \ref{Nomalcoeficients}. When the time modulation $\omega_{\text{s}}$ is slow compared to the frequency of the incident wave ($\omega_{\text{s}} \ll \omega_0$), the modal separation between the two harmonics that carry the main power contribution is large. This can be appreciated in \linebreak Figure \ref{Nomalcoeficients}(a). Conversely, when the time modulation is of the order of the frequency of the incident wave ($\omega_{\text{s}} \sim \omega_0$), then the modal separation between two main harmonics that carry the main power contribution narrows. This is observed in Figures \ref{Nomalcoeficients}(b)-(c). Additionally, it is of interest to note that, in the case $\omega_{\text{s}} = \omega_0$, the harmonic with the highest power contribution is  is $n=-1$ [see Figure \ref{Nomalcoeficients}(c)]. This means that great amount of power could be transferred from the incident wave to $(-1)$-th harmonic, fact that is of potential interest to be exploited in engineering for beamforming purposes and the creation of analog mixers \cite{Taravati2022}. Furthermore, the symmetry of the considered tangential field $E(t)$ causes that higher-order even harmonics are null, fact that is corroborated by the FDTD simulation. It is expected that higher-order even harmonics are no longer null if the time that the screen is in the air state and in the metal state is not the same; that is, if the change occurs at an instant different from $T_\mathrm{s}/2$. Finally, \Fig{N_f_0coma25_final} illustrates the amplitude of Floquet harmonics when $\omega_\textrm{s} = 4\omega_0$. Cases where the time modulation is notably faster than the vibration of the incident wave ($\omega_\textrm{s} \gg \omega_0$) provoke that most of the diffracted power transfer to the fundamental ($n=0$) and $n=-2$ harmonics. According to eq.\eqref{angle}, both harmonics have diffraction angles in the same direction of incidence $\theta_0 = \theta_{-2} = \theta$. Moreover, when the screen commutes fast, it is expected that a significant amount of power is transferred to angles close to the normal. This that will be verified in further FDTD simulations.     

Figure \ref{ObliqueFDTD} illustrates a FDTD simulation showing the electric field distribution when different time modulations $\omega_{\text{s}}$ are considered. In this case, oblique TE incidence ($\theta=30^\mathrm{o}$) is assumed. For a better visualization of the diffraction phenomena, only transmitted waves are plotted in this case. {Black arrows indicate the theoretical propagation direction of the Floquet harmonics that carry most of the diffracted power.} By looking at Figure \ref{ObliqueFDTD}(a), it can observed that, in those cases where $\omega_{\text{s}}$ is much less compared to the frequency of the incident wave ($\omega_{\text{s}} \ll \omega_0$), higher-order harmonics diffract with angles very close to that of the fundamental harmonic $n=0$. As a consequence, they appear to overlap. 
This is in agreement with the theoretical expression for the angles of the transmitted waves \eqref{angle}.  In this scenario, the system acts as a pulsed source of angular frequency $\omega_{\text{s}}$.

Cases where $\omega_{\text{s}} \ll \omega_0$ are not desirable in order to control the steering angle. However, the situation is different when $\omega_{\text{s}} \sim \omega_0$. The fact that the modulation frequency is comparable to the frequency of the incident wave causes that the angle of reflection/transmission of higher-order modes broadens. This is illustrated in Figures \ref{ObliqueFDTD}(b)-(c). Concretely, let us focus on Figure \ref{ObliqueFDTD}(c), where the time modulation is identical to the frequency of the incident wave. The angle of the transmitted higher-order waves $\theta_n$ can be calculated analytically according to eq.\eqref{angle}. The resulting theoretical values are $\theta_0 =30^\mathrm{o}$, $\theta_{1} = 14.48^\mathrm{o}$, $\theta_{2} = 9.59^\mathrm{o}$, and $\theta_{3} = 7.18^\mathrm{o}$. These values are in agreement with the FDTD simulation: $\theta_0^\mathrm{FDTD} =\arctan(11.5/20) = 29.89^\mathrm{o}, \theta_1^\mathrm{FDTD} = \arctan(5.1/20) = 14.31 ^\mathrm{o}, \theta_2^\mathrm{FDTD} =\arctan(3.4/20) = 9.65^\mathrm{o}, \theta_3^\mathrm{FDTD} =\arctan(2.6/20) = 7.41^\mathrm{o}$.

\begin{figure}[!t]
\centering
\hspace{-0.4cm}
\subfigure[\hspace{-1cm}]{
\includegraphics[width=0.234\textwidth]{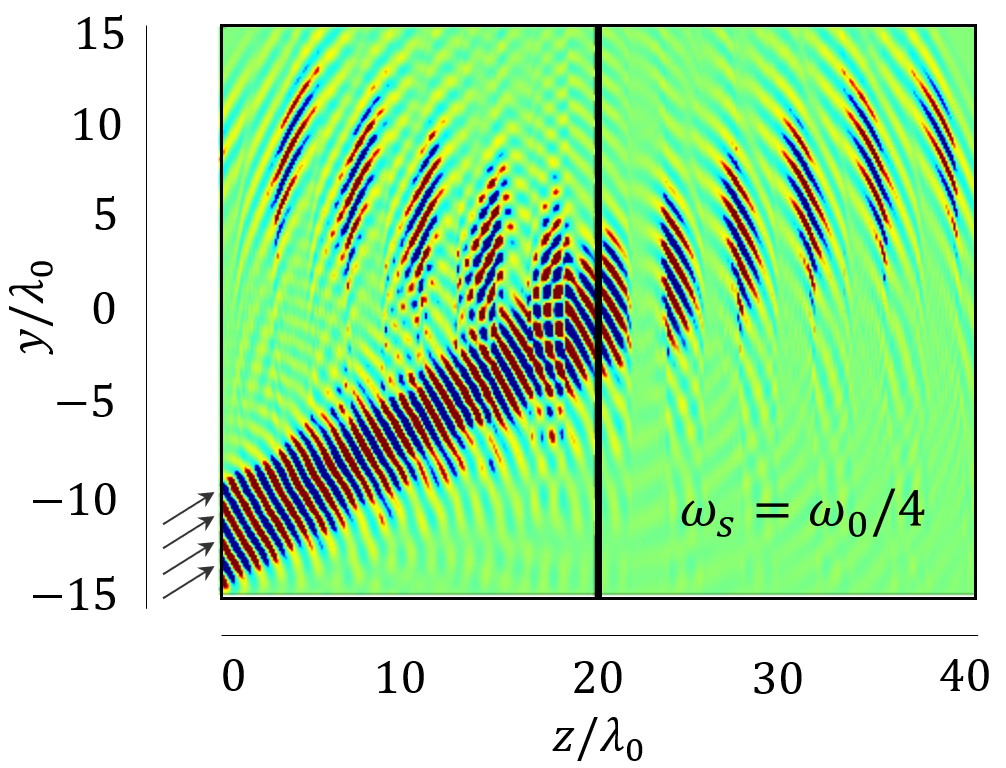}}
\hspace{-0.2cm}
\subfigure[\hspace{-1cm}]{
\includegraphics[width=0.236\textwidth]{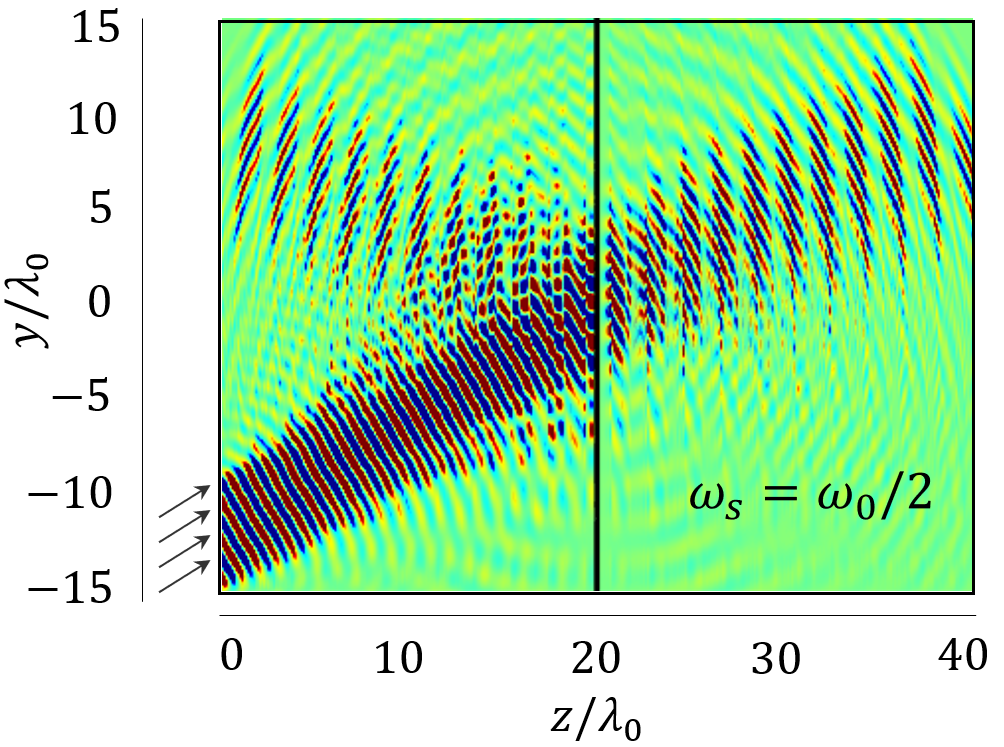}}\\
\hspace{-0.4cm}
\subfigure[\hspace{-1cm}]{
\includegraphics[width=0.236\textwidth]{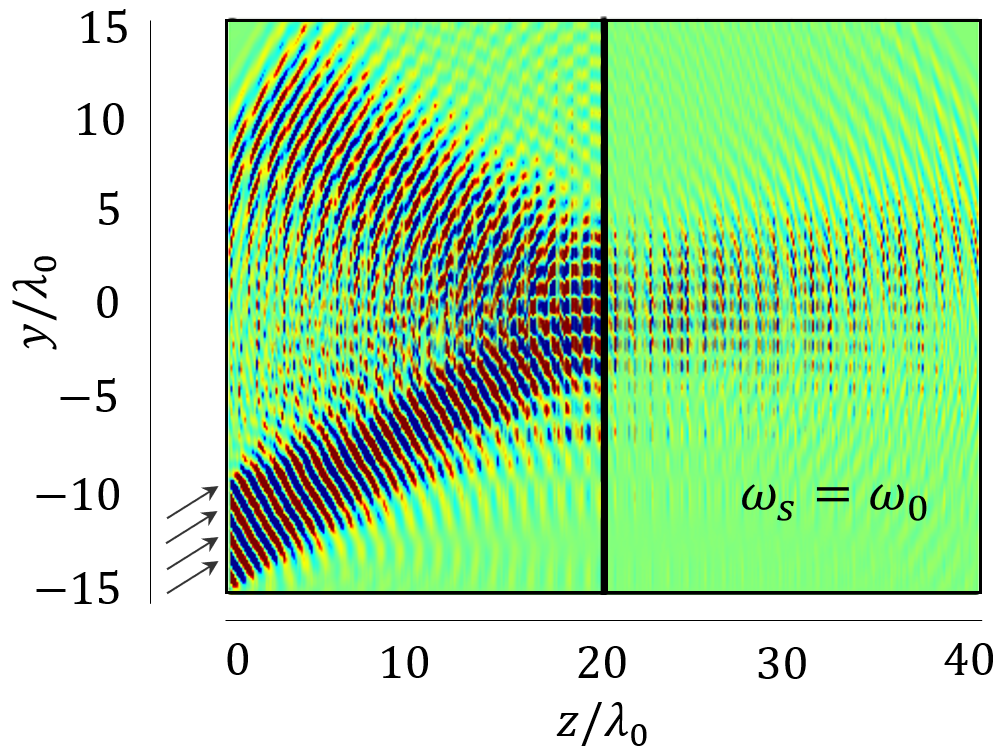}}
\hspace{-0.2cm}
\subfigure[\hspace{-1cm}]{
\includegraphics[width=0.238\textwidth]{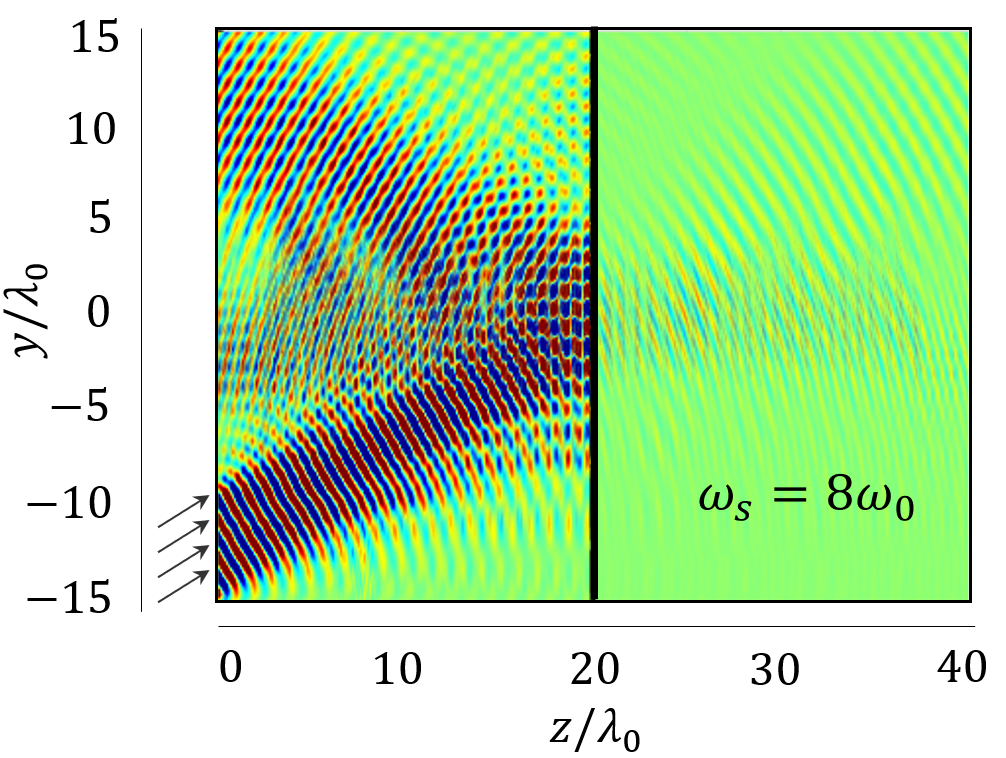}}
\caption{\small {FDTD simulation (electric field) showing reflection and transmission through a time-modulated metallic screen  when a TE-polarized oblique wave ($\theta = 30^\mathrm{o}$) impinges the structure. Cases: (a) $\omega_s = \omega_0/4$, (b) $\omega_s = \omega_0$, (c) $\omega_s = \omega_0$, (d) $\omega_s = 8\omega_0$.} } 
\label{FDTD_RandT}
\end{figure}

As discussed previously, cases $\omega_\textrm{s} \gg \omega_0$ are expected to locate most of the diffracted energy in the direction of incidence ($\theta=30^\mathrm{o}$) or relatively close to it. This can be appreciated in the FDTD simulation shown in Figure \ref{ObliqueFDTD}(d).  This can be  understood by looking at the sampling that the time-varying screen causes to the incident wave [see Figures \ref{time_variation_fast}(a) and \ref{time_variation_fast}(b)]. As the sampling is finer ($\omega_\textrm{s}$ increases), the original incident wave is reproduced in a better way. Therefore, the basis function (field profile) $E(t)$ turns progressively into a discrete version of $\sin(\omega_0t)$, sampled at integer multiples of $T_\textrm{s}$. Thus the most of power is carried by the harmonics $n = 0$ and $n = -2$, which are electromagnetically identical ($|k_{0}| = |k_{-2}|$). The rest of diffraction orders transmits and reflects at angles close to the normal and with lower (generally much lower) amplitudes. Comparison between analytical and FDTD diffracted angles in Figure \ref{ObliqueFDTD}(d) show a good agreement. Analytical values are $\theta_0 =30^\mathrm{o}$, $\theta_{2} = 9.59^\mathrm{o}$ and $\theta_{4} = 5.74^\mathrm{o}$, while values obtain by FDTD are $\theta_0^\mathrm{FDTD} =\arctan(11.3/20) = 29.46^\mathrm{o},  \theta_2^\mathrm{FDTD} = \arctan(3.5/20) = 9.93^\mathrm{o}, \theta_4^\mathrm{FDTD} =\arctan(2/20) = 5.71^\mathrm{o}$. In addition, little transmission observed at angles below $0^\mathrm{o}$ is due to numerical noise and should not be confused with waves actually propagating.

Now, Figure \ref{FDTD_RandT} presents a more general FDTD scenario involving reflected and transmitted waves. In this case, the time-varying screen is located at position $z=20\lambda_0$ and the incident angle is $\theta = 30^\mathrm{o}$. Following the previous discussion, slow time modulations ($\omega_\mathrm{s}  \ll \omega_0$) provoke that the time-varying screen acts as a pulsed source, both in reflection and transmission, with practically null diffraction.  This is sketched in Figures \ref{FDTD_RandT}(a) and \ref{FDTD_RandT}(b). Naturally, the separation between consecutive wavefronts is related to the ratio $\omega_0/\omega_\mathrm{s} $. Modulations of the kind $\omega_\mathrm{s} \sim \omega_0$ [see Figure \ref{FDTD_RandT}(c)]  show the richest pattern in terms of diffraction, while fast time modulations ($\omega_\mathrm{s}  \gg \omega_0$) mainly diffract waves in the specular-reflection and direct-transmission angles as well as in regions near the normal [see Figure \ref{FDTD_RandT}(d)]. Apparently, it is seen in Figures \ref{FDTD_RandT}(a)-(d) that the amplitude of the reflected waves increases (transmission decreases) as $\omega_0 / \omega_\mathrm{s}$ is smaller.

\subsection{Dielectric Media}

\begin{figure}[!t]
\centering
\subfigure[\hspace{-0.6cm}]{
\includegraphics[width=0.15\textwidth]{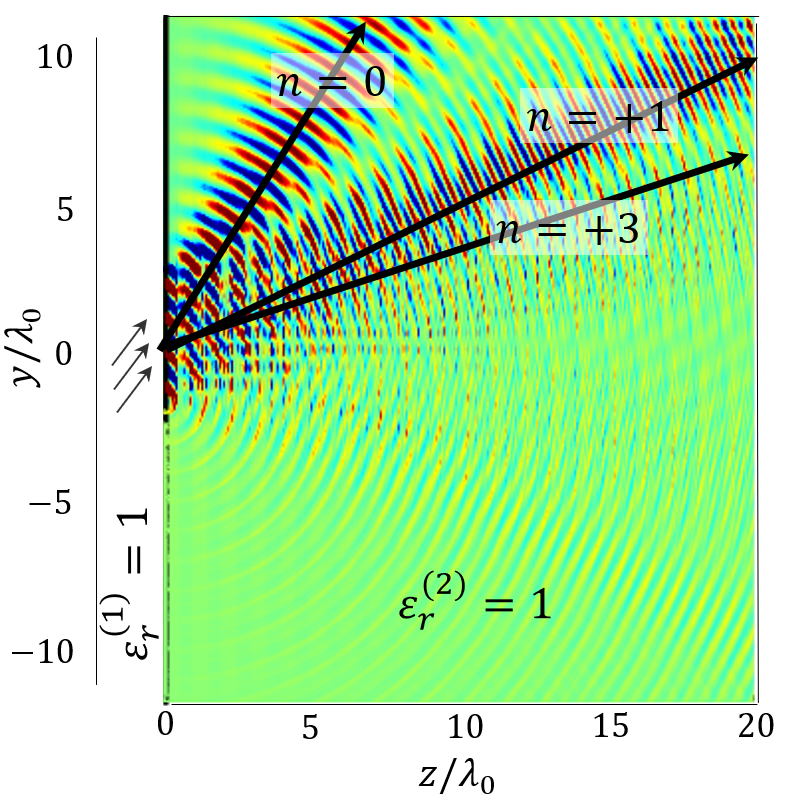}}
\subfigure[\hspace{-0.6cm}]{
\includegraphics[width=0.15\textwidth]{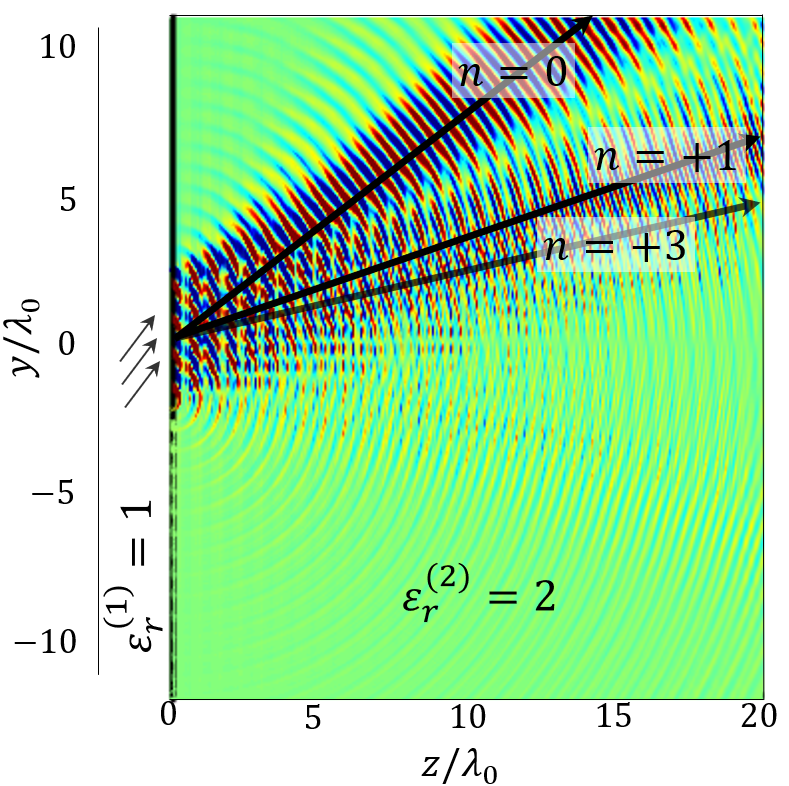}}
\subfigure[\hspace{-0.6cm}]{
\includegraphics[width=0.15\textwidth]{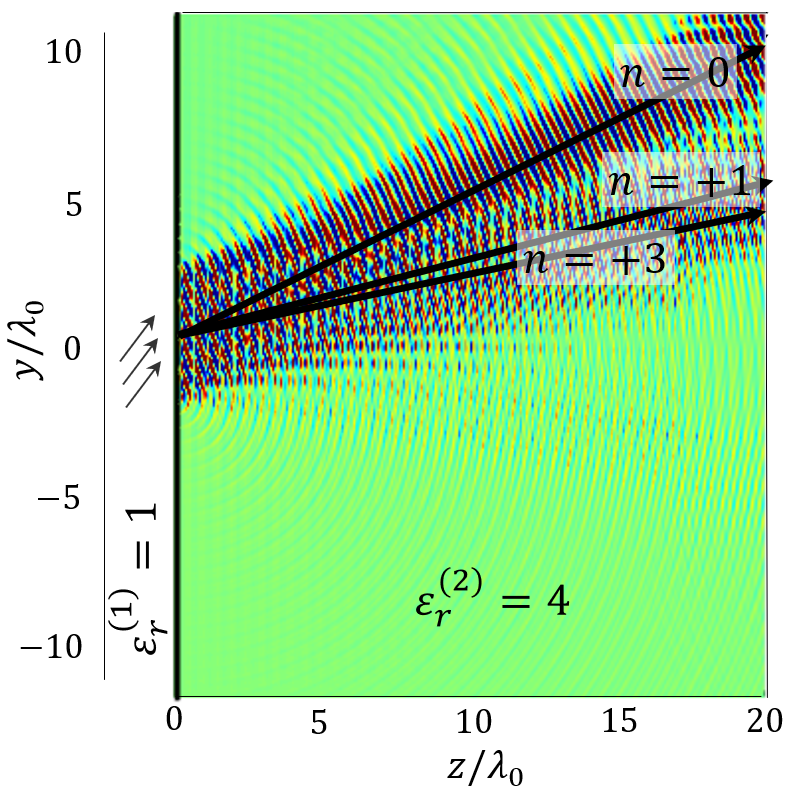}}
	\caption{\small {FDTD simulation (electric field) showing transmission through a time-modulated metallic screen loaded with a semi-infinite dielectric $\varepsilon_r$  when a TE-polarized oblique wave ($\theta = 60^\mathrm{o}$) impinges the structure. Cases: (a) $\varepsilon_r^{(2)}=1$, (b) $\varepsilon_r^{(2)}=2$, (c) $\varepsilon_r^{(2)}=4$. Parameters: $\omega_{\text{s}} = \omega_0$.}} 
\label{Stratified_field}
\end{figure}

Figure \ref{Stratified_field} illustrates a FDTD simulation showing the electric field distribution for a structure formed by a time-varying screen backed by a semi-infinite medium of relative permittivity $\varepsilon_r^{(2)}$.  Input medium is considered to be air and $\omega_s = \omega_0$ to enhance the spacing between angles of different diffraction orders. Then, the angle of transmission of the $n$-th diffracted wave, $\theta_n$, is calculated analytically according to eq. \eqref{angle} and compared to numerical FDTD simulations in Table \ref{table1}.  Results show a good agreement between theory and numerical computations. As already predicted in Sec. II.D, large values of $\varepsilon_r^{(2)}$ provoke that the diffracted waves approach to the normal ($\theta = 0$). If the output medium is dense, higher-order harmonics concentrate on a small angular region close to the normal, which makes them very difficult to visualize in a field chart. As a consequence, diffraction orders $n\geq 3$ are not compared to the FDTD in Table \ref{table1}.

\begin{table}  
\setlength{\tabcolsep}{8pt}
\caption{Diffraction angle $\theta_n$ (deg) for the $n$th-harmonic in a structure formed by a time-varying screen backed by a semi-infinite medium of relative permittivity $\varepsilon_r^{(2)}$. Parameters: $\omega_{\text{s}} = \omega_0$, $\varepsilon_r^{(1)} = \mu_r^{(1)} = \mu_r^{(2)} = 1$, $\theta=60^\mathrm{o}$.}
\vspace{0.2cm}
\centering \label{table1}
\begin{tabular}{ @{} c *4c @{}} 
\toprule
 \multicolumn{1}{c}{Diffraction Angle (deg)}    & $\theta_0$  & $\theta_1$  & $\theta_2 $  & $\theta_3 $  \\ 
\midrule
 Theory $\big(\varepsilon_r^{(2)}=1 \big)$ & $60$ & $25.66$ & $16.78$ & 12.50 \\ 
 FDTD $\big(\varepsilon_r^{(2)}=1 \big)$ &  $59.40$ &  $25.87$ &  $16.69$ & - \\
\midrule
 Theory $\big(\varepsilon_r^{(2)}=2 \big)$ & $37.76$ & $17.83$ & $11.78$ & 8.81 \\ 
 FDTD $\big(\varepsilon_r^{(2)}=2 \big)$ &  $37.48$ &  $18.00$ &  $11.31$ & - \\
 \midrule
 Theory $\big(\varepsilon_r^{(2)}=4 \big)$ & $25.66$ & $12.50$ & $8.30$ & 6.21 \\ 
 FDTD $\big(\varepsilon_r^{(2)}=4 \big)$ &  $26.10$ &  $12.68$ &  $8.53$ & - \\
 \bottomrule
 \end{tabular}
\end{table}

\subsection{Applications}

At the light of the present results, the applications of the proposed time-periodic metallic screen are directly linked to its diffractive and pulsed-source behavior. In general, diffraction gratings (space-only, time-only or spacetime-modulated) are commonly used as filters, monochromators, spectrometers, lasers, wavelength division multiplexing devices, holographers, polarizers, beamformers, direction-of-arrival (DoA) estimators, and in many other microwave, photonic and optical applications. Naturally, gratings can be of reflective or transmissive types, depending on the intended operation. The main advantages of time gratings compared to traditional space gratings are related to the inherent capabilities of time-periodic structures to mix frequencies. This is an inherent property related to the time periodicity, $\omega_n = \omega_0 + n\omega_s$. This fact is a valuable asset that can be exploited in engineering for the design of analog FSS-based mixers.

Furthermore, combining time and space periodicities could directly lead to nonreciprocal responses for the considered devices. The search of nonreciprocity has become a hot topic in electromagnetism in the last years, especially if it is achieved by avoiding bulky magnets or, in general, magnetic materials. Next-generation networks based on beamforming can take advantage of the nonreciprocal response of spacetime-modulated metamaterials. In fact, the present time-varying screen can efficiently act as a beamformer by simply adjusting the frequency modulation $\omega_s$.

On the other hand, pulsed sources are demanded in many fields of physics, engineering and medicine.  For instance, pulsed electromagnetic field therapy (PEMF) uses electromagnetic fields to heal non-union fractures and other injuries [54]. Spectrometry and spectroscopy are traditional applications of pulsed waves [55]. More exotic applications of pulsed sources can be found in agriculture, where pulsed electromagnetic fields stimulate biological effects of chemically active species in the plasma [56].  Actually, time-modulated metamaterials are currently being investigated as a feasible and alternative source of pulsed waves \cite{PulsedSource2022}. In fact, the proposed time-varying metamaterial can act as a pulsed source if the modulation frequency is small compared to the vibration of the incident wave ($\omega_s \ll \omega_0$). The waveform of the generated pulses will coincide with the waveform of the incident wave. In this case, the incident wave is a plane wave of sinusoidal nature, so the pulses are of sinusoidal nature too.

In addition, the proposed structure can act as an analog sampler when the modulation frequency is much greater than the vibration of the incident wave \linebreak ($\omega_s \gg \omega_0$). This phenomenon can be visualized in Figure \ref{time_variation_fast}. {The} modulation frequency $\omega_s$ of the time-periodic screen would control the sampling rate. 

\section{Conclusion}
Previous state-of-art works have mainly focused on the study of dielectric-based metamaterials with spatiotemporal variations. In this work, we have presented an analytical framework that serves as a basis for the study of time-modulated {metal-based} metamaterials. The time periodicity of the problem allows us to expand the electromagnetic fields in terms of a Floquet-Bloch series. By imposing the continuity of the electric field and the instantaneous Poynting vector across the time interface, the Floquet coefficients, dispersion curves, diffraction angles as well as the reflection/transmission coefficients are derived. The present approach comes with an associated equivalent circuit, formed by two input and output transmission lines and an equivalent admittance that models the time-varying screen. This equivalent admittance groups an infinitely set of parallel transformers and transmission lines (one per each harmonic). By checking the dispersion curves of the system, we have seen the linearity (non-dispersion) of the modes and the absence of stopbands. In addition, we have shown that higher-order harmonics are propagative, which is a notable difference compared to spatially-modulated diffraction gratings where higher-order modes are typically evanescent and carry reactive power. Some analytical and numerical (self-implemented FDTD) results are provided in order to validate the approach. Results show that time-modulated metallic screens can act either as pulsed sources (when $\omega_\textrm{s}  \ll \omega_0$) or as beamformers (when $\omega_\textrm{s} \sim \omega_0$) to redirect power. Moreover, cases where the time modulation is notably faster compared to the frequency of the incident wave ($\omega_\textrm{s}  \gg \omega_0$) provoke that most of the diffracted power redirect to the fundamental harmonic and to angles close to the normal. These are interesting features that can be considered for the manipulation and reconfiguration of electromagnetic waves in future wireless communications systems.

\begin{acknowledgments}
This work was supported in part by the Spanish Government under Project PID2020-112545RB-C54 and Project RTI2018-102002-A-I00, in part by ``Junta de Andalucía'' under Project B-TIC-402-UGR18, Project A-TIC-608-UGR20, Project PYC20-RE-012-UGR and Project P18.RT.4830, and and in part by a Leonardo Grant of the BBVA foundation. 
The authors acknowledge the support of the BBVA foundation for the funds associated to a project belonging to
the program Leonardo Grants 2021 for researchers and cultural
creators from the BBVA foundation.
\end{acknowledgments}

\appendix

\section{Higher-order Propagative Waves}
The propagation constant in medium $(i)$ reads
\begin{align}
    \label{beta2} \beta_{n}^{(i)} &= \sqrt{\varepsilon_r^{(i)}\mu_r^{(i)}} \,
    \sqrt{\left[k_{0} + n \frac{\omega_s}{c}\right]^2 - \left[k_{0} \sin (\theta)\right]^2}
\end{align}
In the case of dealing with higher-order waves ($|n|\gg 1$), 
$\left[k_{0} + n \omega_s/c\right]^2 \gg \left[k_{0} \sin (\theta)\right]^2$ and $|n \omega_s/c| \gg k_0$. As a consequence, \eqref{beta2} can be simplified to
\begin{align}
    \label{beta_high} \beta_{n}^{(i)} \approx  \frac{\sqrt{\varepsilon_r^{(i)}\mu_r^{(i)}}}{c} \,
      n \omega_s, \quad |n|\gg 1\, \,.
\end{align}
By inserting eq. \eqref{beta_high} into eqs. \eqref{admittanceTM}, \eqref{admittanceTE} and noticing that $|n \omega_s| \gg \omega_0$ when $|n|\gg 1$, we would reach the expression for the wave admittances of higher-order waves: 
\begin{align}
\label{admittance_high}Y_{n}^{(i),\textrm{TM}} = Y_{n}^{(i),\textrm{TE}} \approx \sqrt{\frac{\varepsilon_r^{(i)} \varepsilon_0}{\mu_r^{(i)} \mu_0}} = Y_0^{(i)}, \quad |n|\gg 1\, \,.
\end{align}
As it can be appreciated, TM and TE admittances are identical, real-valued (in lossless media) and independent from $n$ in the case of considering higher-order waves.

\section{FDTD Simulations}
Numerical simulations are performed with a self-implemented finite-difference time-domain (FDTD) method programmed in Matlab. Our FDTD approach is directly derived from Maxwell's equations by assuming media free of charges, and then particularized to  2-D cases where the incident plane wave is TE polarized ($E_x$, $H_y$, $H_z$). We have worked with a staggered FDTD scheme; namely, $\mathbf{H}^{(\Delta t \cdot n)}$ and $\mathbf{E}^{(\Delta t \cdot n/2)}$, where $\Delta t = C \Delta y / c$ is the time step, $n$ is an integer and $C$ is a dimensionless parameter associated to the CFL stability condition. To suppress numerical reflections, second-order Engquist-Majda absorbing boundary conditions have been employed. For the simulations, we have considered a uniform square grid with spatial resolution \linebreak[1] $\Delta_y = \Delta_z = \lambda_0/35$, and $C=0.4$ (for stability: $C<1/\sqrt{2}$  if $\Delta_y = \Delta_z$). Furthermore, our FDTD approach can include both static [$\varepsilon \equiv \mathrm{ctt}$] and time-modulated dielectrics [$\varepsilon = \varepsilon(t)$]. However,  the time-varying metallic screen, located at $z = z_\mathrm{pos}$, is directly modelled in our case as a time-dependent boundary condition for the electric field \linebreak[5] [$E_x(y, z_\mathrm{pos}, t)= 0$, for $-T_\mathrm{s}/2\leq t < 0,\, \forall y$]. Additionally, Floquet coefficients can be extracted from the FDTD as a part of a post-processing step. To do so, the tangential electric field $E(t)$, evaluated along the time period at the screen's interface $z_\mathrm{pos}$, should be stored and numerically integrated [see eq. \eqref{ordenn}].



\end{document}